\documentclass{article}
\usepackage{subfiles}
\usepackage{times}

\newif\iflong
\newif\ifshort

\longtrue

\iflong
\else
\shorttrue
\fi

\usepackage{amsmath}
\usepackage{amssymb,amsthm}
\usepackage{bm}
\usepackage{tikz}

\usetikzlibrary{patterns}

\usepackage{todonotes} 
\usepackage{xspace} 

\usepackage{paralist}

\newtheorem{theorem}{Theorem}
\newtheorem{lemma}[theorem]{Lemma}

\newtheorem{definition}[theorem]{Definition}

\usepackage{mathproblem}

\newcommand\blfootnote[1]{%
  \begingroup
  \renewcommand\thefootnote{}\footnote{#1}%
  \addtocounter{footnote}{-1}%
  \endgroup
}

\makeatletter
\g@addto@macro\bfseries{\boldmath}
\makeatother

\newcommand{\Nat}{\mathbb{N}}
\newcommand{\Int}{\mathbb{Z}}
\newcommand{\bigoh}{\ensuremath{{\mathcal O}}}

\newcommand{\SB}{\{\,} \newcommand{\SM}{\;{|}\;} \newcommand{\SE}{\,\}}

\newcommand{\cc}[1]{{{\normalfont\textsf{#1}}}\xspace}

\newcommand{\NP}{\cc{NP}}
\newcommand{\FPT}{\cc{FPT}}
\newcommand{\XP}{\cc{XP}}
\newcommand{\Weft}{{\cc{W}}}
\newcommand{\W}[1]{{\Weft}{{[#1]}}}
\newcommand{\yes}{\textsc{Yes}}

\newcommand{\cH}{\mathcal{H}}
\newcommand{\cR}{\mathcal{R}}
\newcommand{\cRc}[1]{\mathcal{R}_{#1}}

\newcommand{\reca}{\alpha}
\newcommand{\recb}{\beta}
\newcommand{\recc}{\gamma}

\newcommand{\pP}{\mathcal{P}}
\newcommand{\pQ}{\mathcal{Q}}

\DeclareMathOperator*{\tw}{\normalfont\textbf{tw}}
\DeclareMathOperator*{\rw}{\normalfont\textbf{rw}}
\newcommand{\classtreewidth}[1]{\bm{{\operatorname{tw}_{#1}}}}

\newcommand{\MSO}{\(\normalfont\text{CMSO}_1\)}

\newtheorem{observation}{Observation}
\newtheorem{fact}{Fact}

\renewcommand{\todo}[1]{}

\ifshort
\title{Measuring what Matters: A Hybrid Approach to Dynamic Programming with Treewidth}
\fi
\iflong
\title{Measuring what Matters: A Hybrid Approach to Dynamic Programming with Treewidth}
\fi

\author{
Eduard Eiben$^1$ \and
Robert Ganian$^2$ \and
Thekla Hamm$^2$ \and
O-joung Kwon$^3$\\
$^1$ University of Bergen, Norway\\
$^2$ Vienna University of Technology, Austria\\
$^3$ Incheon National University, South Korea
}

\begin{document}
\maketitle

\begin{abstract}
We develop a framework for applying treewidth-based dynamic programming on graphs with ``hybrid structure'', i.e., with parts that may not have small treewidth but instead possess other structural properties. Informally, this is achieved by defining a refinement of treewidth which only considers parts of the graph that do not belong to a pre-specified tractable graph class. Our approach allows us to not only generalize existing fixed-parameter algorithms exploiting treewidth, but also fixed-parameter algorithms which use the size of a modulator as their parameter. As the flagship application of our framework, we obtain a parameter that combines treewidth and rank-width to obtain fixed-parameter algorithms for \textsc{Chromatic Number}, \textsc{Hamiltonian Cycle}, and~\textsc{Max-Cut}.
\end{abstract}

\section{Introduction}
\label{sec:intro}

Over the past decades, the use of structural properties of graphs to obtain efficient algorithms for \NP-hard computational problems has become a prominent research direction in computer science. Perhaps the best known example of a structural property that can be exploited in this way is the tree-likeness of the inputs, formalized in terms of the decomposition-based structural parameter \emph{treewidth}\iflong~\cite{RobertsonS84}\fi. It is now well known that a vast range of fundamental problems admit so-called \emph{fixed-parameter} algorithms parameterized by the treewidth of the input graph -- that is, can be solved in time $f(k)\cdot n^{\bigoh(1)}$ on $n$-vertex graphs of treewidth $k$ (for some computable function $f$). We say that such problems are \FPT parameterized by treewidth.

On the other hand, dense graphs are known to have high treewidth and hence require the use of different structural parameters; the classical example of such a parameter tailored to dense graphs is \emph{clique-width}~\cite{CourcelleOlariu00}. Clique-width is asymptotically equivalent to the structural parameter \emph{rank-width}~\cite{OumSeymour06}, which is nowadays often used instead of clique-width due to a number of advantages (rank-width is much easier to compute~\cite{HlinenyOum08} and can be used to design more efficient fixed-parameter algorithms than clique-width~\cite{GanianHlineny10,GanianHO13}). While rank-width (or, equivalently, clique-width) dominates\footnote{Parameter $\alpha$ dominates parameter $\beta$ if for each graph class with bounded $\beta$, $\alpha$ is also bounded.} treewidth and can be used to ``lift'' fixed-parameter algorithms designed for treewidth to well-structured dense graphs for a number of problems, there are also important problems which are \FPT parameterized by treewidth but \W1-hard (and hence probably not~\FPT) parameterized by rank-width. The most prominent examples of such problems are \textsc{Chromatic Number}~\cite{FominGLS10}, \textsc{Hamiltonian Cycle}~\cite{FominGLS10}, and~\textsc{Max-Cut}~\cite{FominGLS14}.


Another generic type of structure used in algorithmic design is based on measuring the size of a \emph{modulator} (i.e., a vertex deletion set)~\iflong\cite{Cai03}\fi to a certain graph class. Basic examples of parameters based on modulators include the vertex cover number (a modulator to edgeless graphs)~\ifshort\cite{FellowsLMRS08} \fi\iflong\cite{FellowsLMRS08,FominJansenP14} \fi and the feedback vertex set number (a modulator to forests)\ifshort~\cite{BodlaenderJansenKratsch13b}\fi\iflong~\cite{BodlaenderJansenKratsch13b,JansenBodlaender13}\fi. For dense graphs, modulators to graphs of rank-width $1$ have been studied~\cite{EibenGK18,KimK17}, and it is known that for every constant $c$ one can find a modulator of size at most $k$ to graphs of rank-width $c$ (if such a modulator exists) in time $f(k)\cdot n$~\cite{KanteKKP17}.
 However, the algorithmic applications of such modulators have remained largely unexplored up to this point.

\smallskip
\noindent \textbf{Our Contribution.}
We develop a class of \emph{hybrid parameters} which combines the foremost advantages of treewidth and modulators to obtain a ``best-of-both-worlds'' outcome. In particular, instead of measuring the treewidth of the graph itself or the size of a modulator to a graph class $\cH$, we consider the treewidth of a (torso of a) modulator to $\cH$\iflong\footnote{Formal definitions are provided in Section~\ref{sec:H-tw}}\fi. This parameter, which we simply call $\cH$-treewidth, allows us to lift previously established tractability results for a vast number of problems from treewidth and modulators to a strictly more general setting. As our first technical contribution, we substantiate this claim with a meta-theorem that formalizes generic conditions under which a treewidth-based algorithm can be generalized to $\cH$-treewidth; the main technical tool for the proof is an adaptation of \emph{protrusion replacement} techniques~\cite[Section~4]{BodlaenderFominLokshtanovPenninkxSaurabhThilikos16}.

As the flagship application of $\cH$-treewidth, we study the case where $\cH$ is the class $\cRc{c}$ of graphs of rank-width at most $c$ (an arbitrary constant). $\cRc{c}$-treewidth hence represents a way of lifting treewidth towards dense graphs that lies ``between'' treewidth and rank-width. We note that this class of parameters naturally incorporates a certain scaling trade-off: $\cRc{c}$-treewidth dominates $\cRc{c-1}$-treewidth for each constant $c$, but the runtime bounds for algorithms using $\cRc{c}$-treewidth are worse than those for $\cRc{c-1}$-treewidth. 

Our first result for $\cRc{c}$-treewidth is a fixed-parameter algorithm for computing the parameter itself. We then develop fixed-parameter algorithms for \textsc{Chromatic Number}, \textsc{Hamiltonian Cycle} and \textsc{Max-Cut} parameterized by $\cRc{c}$-treewidth; moreover, in $2$ out of these $3$ cases the parameter dependencies of our algorithms are essentially tight. These algorithms represent generalizations of:

\begin{enumerate}
\vspace{-0.3cm}
\item classical fixed-parameter algorithms parameterized by treewidth~\cite{DowneyFellows13},
\item polynomial-time algorithms on graphs of bounded rank-width~\cite{GanianHO13}, and
\item (not previously known) fixed-parameter algorithms parameterized by modulators to graph classes of bounded rank-width. 
\vspace{-0.3cm}
\end{enumerate}

The main challenge for all of these problems lies in dealing with the fact that some parts of the graph need to be handled using rank-width based techniques, while for others we use treewidth-based dynamic programming. We separate these parts from each other using the notion of \emph{nice $\cH$-tree decompositions}. 
The algorithm then relies on enhancing the known dynamic programming approach for solving the problem on treewidth with a subroutine that not only solves the problem on the part of the graph outside of the modulator, but also serves as an interface by supplying appropriate records to the treewidth-based dynamic programming part of the algorithm. At its core, each of these subroutines boils down to solving an ``extended'' version of the original problem parameterized by the size of a modulator to $\cRc{c}$; in particular, each subroutine immediately implies a fixed-parameter algorithm for the respective problem when parameterized by a modulator to constant rank-width. To give a specific example for such a subroutine, in the case of \textsc{Chromatic Number} one needs to solve the problem parameterized by a modulator to $\cRc{c}$ where the modulator is furthermore precolored.

To avoid any doubt, we make it explicitly clear that the runtime of all of our algorithms utilizing $\cRc{c}$-treewidth has a polynomial dependency on the input where the degree of this polynomial depends on $c$ (as is necessitated by the \W{1}-hardness of the studied problems parameterized by rank-width).

\smallskip
\noindent \textbf{Related Work.}
Previous works have used a combination of treewidth with \emph{backdoors}, a notion that is closely related to modulators, in order to solve 
non-graph problems such as \textsc{Constraint Satisfaction}~\cite{GanianRS17}, \textsc{Boolean Satisfiability}~\cite{GanianRS17b} and \textsc{Integer Programming}~\cite{GanianOR17}. Interestingly, the main technical challenge in all of these papers is the problem of computing the parameter, while using the parameter to solve the problem is straightforward. In the algorithmic results presented in this contribution, the situation is completely reversed: the main technical challenge lies in developing the algorithms (and most notably the subroutines) for solving our targeted problems. Moreover, while the aforementioned three papers focus on solving a single problem, here we aim at identifying and exploiting structural properties that can be used to solve a wide variety of graph problems.
\ifshort

Other parameters which target inputs with hybrid structure include \emph{sm-width}~\cite{SaetherTelle14} and well-structured modulators~\cite{EibenGS18b}. It is not difficult to show that these are different (both conceptually and factually) from $\cRc{c}$-treewidth.  $(\clubsuit)$ \blfootnote{Statements where proofs or details are provided in the appendix are marked with $\clubsuit$.}
\fi
\iflong

Telle and Saether~\cite{SaetherTelle14} proposed a generalization of treewidth towards dense graphs that is based on explicitly allowing a specific operation (graph splits). The resulting parameter is not related to modulators, and we show that it is incomparable to $\cRc{c}$-treewidth.
A subset of the authors previously studied modulator-based parameters that used graph splits and rank-width. The resulting \emph{well-structured modulators} could either dominate rank-width~\cite{EibenGS18b}, or (depending on the specific constants used) be used to obtain polynomial kernels for a variety of problems~\cite{EibenGS18}. Since $\cRc{c}$-treewidth lies ``between'' treewidth and rank-width, it is clear that our class of parameters is different from these previously studied ones. 
\fi


\section{Preliminaries}
\label{sec:prelim}
For $i\in \Nat$, let $[i]$ denote the set $\{1,\dots,i\}$.
All graphs in this paper are simple and undirected.
We refer to the standard textbook~\cite{Diestel00} for basic graph terminology.
\ifshort
For $S\subseteq V(G)$, let $G[S]$ denote the subgraph of $G$ induced by $S$.
For $v\in V(G)$, the set of neighbors of $v$ in $G$ is denoted by $N_G(v)$ (or $N(v)$ when $G$ is clear from the context).
For $A\subseteq V(G)$, let $N_G(A)$ denote the set of vertices in $G-A$ that have a neighbor in $A$.
For a vertex set $A$, an \emph{$A$-path} is a path whose endpoints are contained in $A$ and all the internal vertices are contained in $G-A$.
\fi
\iflong

For a graph $G$, let $V(G)$ and $E(G)$ denote the vertex set and the edge set of $G$, respectively. 
For $S\subseteq V(G)$, let $G[S]$ denote the subgraph of $G$ induced by $S$; if $G$ contains no edges with precisely one endpoint in $S$, then we call $S$ a \emph{connected component} of $G$. For $v\in V(G)$ and $S\subseteq V(G)$, let $G- v$ be the graph obtained from $G$ by removing $v$, and let $G-S$ be the graph obtained by removing all vertices in $S$. 
For $v\in V(G)$, the set of neighbors of $v$ in $G$ is denoted by $N_G(v)$ (or $N(v)$ when $G$ is clear from the context).
For $A\subseteq V(G)$, let $N_G(A)$ denote the set of vertices in $G-A$ that have a neighbor in $A$.
For $v\in V(G)$ and a subgraph $H$ of $G-v$, we say $v$ is \emph{adjacent} to $H$ if it has a neighbor in $H$. 
For a vertex set $A$ of a graph $G$, an \emph{$A$-path} is a path whose endpoints are contained in $A$ and all the internal vertices are contained in $G-A$.

\fi
A set $M$ of vertices in a graph $G$ is called a \emph{modulator} to a graph class $\cH$ if $G-M\in \cH$.
The operation of \emph{collapsing} a vertex set
$X$, denoted $G \circ X$, deletes $X$ from the graph and adds an
edge between vertices $u,v\in V(G-X)$ if $uv\notin E(G)$ and there is an $u$-$v$ path with all internal vertices in $G[X]$.
\ifshort
We assume that the reader is familiar with \emph{parameterized complexity}~\cite{CyganFominKowalikLokshtanovMarxPilipczukPilipcukSaurabh13,DowneyFellows13},
notably with notions such as \FPT, \W{1}, treewidth and Courcelle's Theorem.

\fi
\iflong
\subsection{Parameterized Complexity}

A \emph{parameterized problem} $\pP$ is a subset of $\Sigma^* \times \Nat$ for some finite alphabet $\Sigma$. Let $L\subseteq \Sigma^*$ be a classical decision problem for a finite alphabet, and let $p$ be a non-negative integer-valued function defined on $\Sigma^*$. Then $L$ \emph{parameterized by} $p$ denotes the parameterized problem $\SB(x,p(x)) \SM x\in L \SE$ where $x\in \Sigma^*$.  For a problem instance $(x,k) \in \Sigma^* \times \Nat$ we call $x$ the main part and $k$ the parameter.  
A parameterized problem $\pP$ is \emph{fixed-parameter tractable} (\FPT
in short) if a given instance $(x, k)$ can be solved in time
$f(k) \cdot |x|^{\bigoh(1)}$ where $f$ is an arbitrary computable
function of $k$; we call algorithms
running in this time \emph{fixed-parameter algorithms}.
Similarly, a parameterized problem $\pP$ is in the class \XP\ 
if a given instance $(x, k)$ can be solved in time
$|x|^{f(k)}$ where $f$ is an arbitrary computable
function of $k$, and we call algorithms running in this time \emph{XP algorithms}.

Parameterized complexity classes are defined with respect to {\em
  fpt-reducibility}. A parameterized problem $P$ is {\em
  fpt-reducible} to $Q$ if in time $f(k)\cdot |x|^{O(1)}$, one can
transform an instance $(x,k)$ of $\pP$ into an instance $(x',k')$ of
$\pQ$ such that $(x,k)\in \pP$ if and only if $(x',k')\in \pQ$, and
$k'\leq g(k)$, where $f$ and $g$ are computable functions depending
only on $k$.
Central to parameterized complexity is the following hierarchy of complexity classes, defined by the closure of canonical problems under fpt-reductions:
$\FPT \subseteq \W{1} \subseteq \W{2} \subseteq \cdots \subseteq \XP.$ All inclusions are believed to be strict. In particular, $\FPT \neq \W{1}$ under the Exponential Time Hypothesis~\cite{ImpagliazzoPaturiZane01}.

The class $\W{1}$ is the analog of $\NP$ in parameterized complexity. A major goal in parameterized complexity is to distinguish between parameterized problems which are in $\FPT$ and those which are $\W{1}$-hard, i.e., those to which every problem in $\W{1}$ is fpt-reducible.

We refer the reader to the respective books~\cite{CyganFominKowalikLokshtanovMarxPilipczukPilipcukSaurabh13,DowneyFellows13,Niedermeier06} for more details on parameterized complexity.
Finally, we recall that a parameter $\alpha$ \emph{dominates} a parameter $\beta$ if for each graph class where $\beta$ is bounded, $\alpha$ is also bounded. Two parameters are \emph{incomparable} if neither dominates the other.

\subsection{Treewidth}\label{sec:tw}

A \emph{tree decomposition} of a graph $G=(V,E)$ is a pair 
$(T,\{B_t \mid t\in V(T)\})$
where $B_t \subseteq V$ for every $t \in V(T)$ and $T$ is a tree such that:
\begin{enumerate}
  \item for each edge $\{u,v\}\in E$, there is a $t\in V(T)$ such that $\{u,v\} 
\subseteq B_t$, and \label{twone}
\item for each vertex $v\in V$, $T[\SB t\in V(T) \SM v\in B_t \SE]$ is a non-empty (connected) tree. \label{twtwo}
\end{enumerate}
The \emph{width} of a tree decomposition is $\max_{t \in V(T)} |B_t|-1$.
The \emph{treewidth}~\cite{Kloks94} of $G$
is the minimum width taken over all tree decompositions
of $G$ and it is denoted by $\tw(G)$. We call the elements of $V(T)$ \emph{nodes} and $B_t$ \emph{bags}.


\begin{fact}[\cite{Bodlaender96}]
\label{fact:findtw}
There exists an algorithm which, given an $n$-vertex graph $G$ and an integer $k$,
runs in time $k^{\bigoh(k^3)}\cdot n$, and either
outputs a tree decomposition of $G$ of width at most $k$ or correctly
determines that $\tw(G)>k$. 
\end{fact}

A tree decomposition $(T,\{B_t \mid t \in V(T)\})$ of a graph~$G$ is
\emph{nice} if \(T\) can be rooted such that the following conditions hold:
\begin{enumerate}
	\item Every node of \(T\) has at most two children.
	\item If a node \(t\) of \(T\) has two children \(t_1\) and \(t_2\), then \(B_t = B_{t_1} = B_{t_2}\);
	in this case we call \(t\) a \emph{join node}.
	\item If a node \(t\) of \(T\) has exactly one child \(t'\), then either of the following holds:
	\begin{enumerate}
		\item \(\vert B_t \vert = \vert B_{t'} \vert + 1\), in which case we call \(t\) an \emph{introduce node}, or
		\item \(\vert B_t \vert = \vert B_{t'} \vert - 1\) in which case we call \(t\) a \emph{forget node}
	\end{enumerate}
	\item If a node \(t\) is a leaf, then \(\vert B_t \vert = 1\); in this case we call \(t\) a \emph{leaf node}.
\end{enumerate}

The advantage of nice tree decompositions is that they allow the design of much more transparent dynamic programming algorithms, since one only needs to deal with four specific types of nodes. It is well known (and easy to see) that given a
tree decomposition of a graph $G=(V,E)$ of width at most $k$ and with $\mathcal{O}(|V|)$
nodes, one can construct in linear time a nice
tree decomposition of $G$ with $\mathcal{O}(|V|)$ nodes and width at
most~$k$~\cite{BodlaenderKloks96}. Given a node $t$ in $T$, we let $Y_t$ be the set of all vertices contained in the bags of the subtree rooted at $t$, i.e., $Y_t=B_t\cup \bigcup_{p \text{ is separated from the root by $t$}}B_p$.

\subsection{Rank-width}
\fi
\ifshort \smallskip
\noindent \textbf{Rank-width.}
\fi
For a graph $G$ and $U,W\subseteq V(G)$, let $\bm{A}_G[U,W]$ denote the
$U\times W$-submatrix of the adjacency matrix over the two-element
field $\mathrm{GF}(2)$, i.e., the entry $a_{u,w}$, $u\in U$ and $w\in
W$, of $\bm{A}_G[U,W]$ is $1$ if and only if $\{u,w\}$ is an edge
of~$G$.  The {\em cut-rank} function $\rho_G$ of a graph $G$ is
defined as follows: For a bipartition $(U,W)$ of the vertex
set~$V(G)$, $\rho_G(U)=\rho_G(W)$ equals the rank of $\bm{ A}_G[U,W]$. 

A \emph{rank-decomposition} of a graph $G$ is a pair $(T,\mu)$
where $T$ is a tree of maximum degree 3 and $\mu:V(G)\rightarrow \{t \mid t
\text{ is a leaf of } T\}$ is a bijective function\iflong (See Figure~\ref{fig:rdecC5})\fi. For an edge~$e$ of~$T$, the connected components of $T - e$ induce a
bipartition $(X,Y)$ of the set of leaves of~$T$.  The \emph{width} of
an edge $e$ of a rank-decomposition $(T,\mu)$ is $\rho_G(\mu^{-1} (X))$.
The \emph{width} of $(T,\mu)$ is the maximum width over all edges of~$T$.  The \emph{rank-width} of $G$, $\rw(G)$ in short, is the minimum width over all
rank-decompositions of $G$. We denote by $\cR_i$ the class of all graphs of rank-width at most $i$. 
A \emph{rooted} rank-decomposition is obtained from a rank-decomposition by subdividing an arbitrarily chosen edge, and the newly created vertex is called the \emph{root}.


\iflong
\newenvironment{psmallmatrix}
{\left(\begin{smallmatrix}}
	{\end{smallmatrix}\right)}

\begin{figure}[ht]
\vspace{-0.7cm}
\hspace{0.4cm}
\centering
\tikzstyle{every circle node}=[circle,draw,inner sep=1.0pt, fill=black]
\begin{tikzpicture}[scale=0.9]
\begin{scope}[xshift=-7cm]
\draw
(90+4*72:1) node[circle] (d)  {} node[right=2pt] {$\strut d$} 
(90+0*72:1) node[circle] (c)  {} node[above] {$\strut c$} 
(90+1*72:1) node[circle] (b)  {} node[left=2pt] {$\strut b$} 
(90+2*72:1) node[circle] (a)  {} node[below] {$\strut a$} 
(90+3*72:1) node[circle] (e)  {} node[below] {$\strut e$} 
(a)--(b)--(c)--(d)--(e)--(a)
;
\end{scope}

\draw 
(-3,1)   node[circle] (d) {} node[above] {$\strut d$} 
(-3,-1)  node[circle] (e) {} node[below] {$\strut e$} 
(-2,0)   node[circle] (U) {} 
(0,0)    node[circle] (V) {} 
(0,-1.2)    node[circle] (X) {} node[below] {$\strut a$} 
(2,0)    node[circle] (W) {}
(3,1)   node[circle] (c) {} node[above] {$\strut c$} 
(3,-1)  node[circle] (b) {} node[below] {$\strut b$} 

(d) to node [auto,near start, swap] {$\begin{psmallmatrix}
	0&0&1&1
	\end{psmallmatrix}$} (U)
(e) to node [auto, near start] {$\begin{psmallmatrix}
	1&0&0&1
	\end{psmallmatrix}$} (U)
(U) to node [auto] {$\begin{psmallmatrix}
	0&0&1\\
	1&0&0  
	\end{psmallmatrix}$} (V)
(V) to node [auto] {$\begin{psmallmatrix}
	1&0\\
	0&1\\
	0&0
	\end{psmallmatrix}$} (W)
(V)  to node [auto] {$\begin{psmallmatrix}
	1\\
	0\\
	0\\
	1
	\end{psmallmatrix}$} (X)
(c)--(W)--(b)
(3.42,0.65) node {$\begin{psmallmatrix}
	0\\
	1\\
	1\\
	0
	\end{psmallmatrix}$} 
(3.42,-0.65) node {$\begin{psmallmatrix}
	1\\
	1\\
	0\\
	0
	\end{psmallmatrix}$} 
;
\end{tikzpicture}\vspace{-0.5cm}
\caption{A rank-decomposition of the graph cycle \(C_5\).\vspace{-0.5cm}}
\label{fig:rdecC5}
\end{figure}
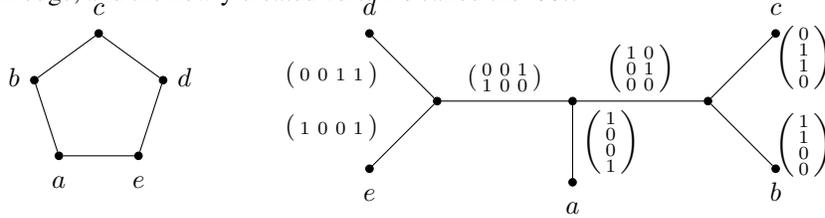
\fi

Unlike clique-width, rank-width can be computed exactly by a fixed-parameter algorithm (which also outputs a corresponding rank-decomposition)\ifshort~\cite{HlinenyOum08}\fi.

\iflong
\begin{theorem}[{\cite[Theorem~7.3]{HlinenyOum08}}]\label{thm:rankdecomp} Let $k \in \Nat$ be a constant and
  $n \geq 2$. For an $n$-vertex graph $G$, we can output a
  rank-decomposition of width at most $k$ or confirm that the
  rank-width of $G$ is larger than $k$ in time $f(k)\cdot n^3$, where $f$ is a computable function.
\end{theorem}
\fi

\iflong
\subsection{Monadic Second-Order Logic}
\label{sub:mso}
\fi
\ifshort
\smallskip \noindent \textbf{Monadic Second-Order Logic.} \fi
Counting Monadic Second-Order Logic (\MSO) is a basic tool to express properties of vertex sets in graphs. The syntax of \MSO\ includes logical connectives $\wedge, \vee, \neg, \Leftrightarrow, \Rightarrow$, variables for vertices and vertex sets, quantifiers $\exists,\forall$ over these variables, and the relations
$a\in A$ where $a$ is a vertex variable and $A$ is a vertex set variable;
$\text{adj}(a,b)$, where $a$ and $b$ are vertex variables and the interpretation is that $a$ and $b$ are adjacent;
equality of variables representing vertices and sets of vertices;
$\text{Parity}(A)$, where $A$ is a vertex set variable and the interpretation is that $|A|$ is even.

The \MSO\ Optimization problem is defined as follows:

\begin{center}
\begin{mathproblem}{\MSO-OPT}
	Instance: & A graph \(G\), a \MSO\ formula \(\phi(A)\) with a free set variable \(A\), and opt \(\in\{\min,\max\}\).\\
	Task: & Find an interpretation of the set \(A\) in \(G\) such that \(G\) models \(\phi(A)\) and \(A\) is of minimum/maximum (depending on opt) cardinality.\\
\end{mathproblem}
\end{center}
  
  From the fixed-parameter tractability of computing rank-width~\cite{HlinenyOum08}, the equivalence of rank-width and clique-width~\cite{OumSeymour06} and Courcelle's Theorem for graphs of bounded clique-width~\cite{CourcelleMakowskyRotics00}
  \iflong (see also later work that establishes the result directly for rank-width~\cite{GanianHlineny10}.)\fi
  it follows that:

\begin{fact}[\cite{GanianHlineny10}]
\label{fact:mso}
\textsc{\MSO-OPT} is \FPT parameterized by $\rw(G)+|\phi|$, where $G$ is the input graph and $\phi$ is the \MSO\ formula.
\end{fact}

%

\iflong
We refer the reader to the books \cite{CourcelleEngelfriet12,FlumGrohe06} for an in-depth overview of Monadic Second Order logic.
\fi

\section{$\cH$-Treewidth} \label{sec:H-tw}

The aim of $\cH$-treewidth is to capture the treewidth of a modulator to the graph class $\cH$. However, one cannot expect to obtain a parameter with reasonable algorithmic applications by simply measuring the treewidth of the graph \emph{induced} by a modulator to $\cH$ -- instead, one needs to measure the treewidth of a so-called \emph{torso}, which adds edges to track how the vertices in the modulator interact through $\cH$. To substantiate this, we observe that \textsc{Hamiltonian Cycle} would become \NP-hard even on graphs with a modulator that (1) induces an edgeless graph, and (2) is a modulator to an edgeless graph, and where (3) each connected component outside the modulator has boundedly many neighbors in the modulator\ifshort~\cite{AkiyamaNishizekiSaito80}~($\clubsuit$)\fi.
\iflong\begin{observation}
	\textsc{Hamiltonian Cycle} remains \NP-hard if we allow the input to consist, not only of a graph \(G\), but also a modulator \(X \subseteq V(G)\) to an edgeless graph and for each connected component \(C\) of \(G[X]\), \(\vert N(C) \vert \leq 3\).
\end{observation}
\begin{proof}
	\textsc{Hamiltonian Cycle} is known to be \NP-hard on cubic bipartite graphs \cite{AkiyamaNishizekiSaito80}.
	Setting \(X\) to be one side of the bipartition of any graph in this graph class is easily seen to satisfy the given conditions.
\end{proof}
\fi

The notion of a \emph{torso} has previously been algorithmically exploited in other settings~\cite{GanianOR17,GanianRS17,MarxW14}, and its adaptation is our first step towards the definition of $\cH$-treewidth (see also Figure~\ref{fig:htw}). 


\begin{definition}[\(\cH\)-Torso]
	Let \(G\) be a graph and \(X\subseteq V(G)\).
	For a graph class \(\cH\), 
	\(G\circ X\) is an \emph{\(\cH\)-torso} of \(G\)
	if each connected component \(C\) of \(G[X]\) satisfies \(C\in \cH\). 
\end{definition}

	\newsavebox\Xa
	\begin{lrbox}{\Xa}
		\begin{tikzpicture}
		\draw[pattern=north east lines] plot[smooth cycle,tension=1] coordinates{(-2.3,-0.5) (-2.6,0) (-2,0.5) (-1.6,0)};
		\end{tikzpicture}
	\end{lrbox}

	\newsavebox\Xb
	\begin{lrbox}{\Xb}
		\begin{tikzpicture}
		\draw[pattern=north east lines] plot[smooth cycle,tension=1] coordinates{(-0.3,-1.5) (0.7,-1.3) (-0.25,-0.8) (-1.55,-1.1)};
		\end{tikzpicture}
	\end{lrbox}

	\newsavebox\Xc
	\begin{lrbox}{\Xc}
		\begin{tikzpicture}
		\draw[pattern=north east lines] plot[smooth cycle,tension=1] coordinates{(3.3,-1.3) (3.9,-1.7) (3.3,-2.1) (2.6,-1.7)};
		\end{tikzpicture}
	\end{lrbox}

	\newsavebox\Xd
	\begin{lrbox}{\Xd}
		\begin{tikzpicture}
		\draw[pattern=north east lines] plot[smooth cycle,tension=1] coordinates{(2.6,1.3) (3.35,1.7) (2.9,2) (1.75,1.7)};
		\end{tikzpicture}
	\end{lrbox}

	\newsavebox\X
	\begin{lrbox}{\X}
		\begin{tikzpicture}
		\draw[pattern=north east lines] plot[smooth cycle,tension=1] coordinates{(3.3,-0) (4.5,-1.7) (3.3,-3) (2.6,-1.7)};
		\end{tikzpicture}
	\end{lrbox}

	\tikzstyle{vertex}=[circle, fill, inner sep=1pt]
	
	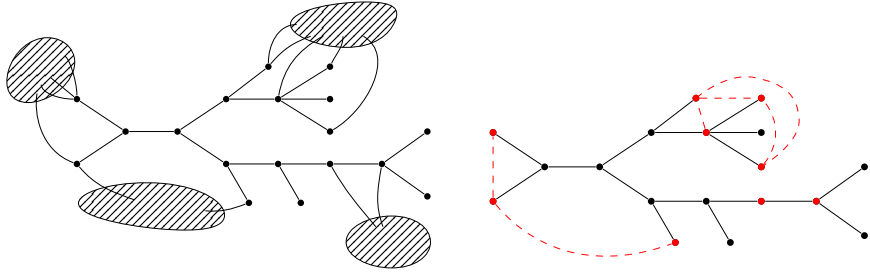
\begin{figure} 
		\centering
		\begin{minipage}{0.475\textwidth}{
			\resizebox{\textwidth}{!}{
				\begin{tikzpicture}
				\useasboundingbox (-2.6,-2.6) rectangle (3.9,2);
				
				\node[vertex] (r) at (0,0) {};
				\node[vertex] (v1) at (-0.8,0) {};
				\node[vertex] (v11) at (-1.55,-0.5) {};
				\node[vertex] (v12) at (-1.55,0.5) {};
				\node[vertex] (v2) at (0.75,-0.5) {};
				\node[vertex] (v21) at (1.1,-1.1) {};
				\node[vertex] (v22) at (1.55,-0.5) {};
				\node[vertex] (v221) at (1.9,-1.1) {};
				\node[vertex] (v222) at (2.35,-0.5) {};
				\node[vertex] (v2221) at (3.15,-0.5) {};
				\node[vertex] (v22211) at (3.85,0) {};
				\node[vertex] (v22212) at (3.85,-1) {};
				\node[vertex] (v3) at (0.75,0.5) {};
				\node[vertex] (v31) at (1.55,0.5) {};
				\node[vertex] (v311) at (2.35,0) {};
				\node[vertex] (v312) at (2.35,0.5) {};
				\node[vertex] (v313) at (2.35,1) {};
				\node[vertex] (v32) at (1.4,1) {};
				\draw (r)--(v1);
				\draw (v1)--(v11);
				\draw (v1)--(v12);
				\draw (r)--(v2);
				\draw (v2)--(v21);
				\draw (v2)--(v22);
				\draw (v22)--(v221);
				\draw (v22)--(v222);
				\draw (v222)--(v2221);
				\draw (v2221)--(v22211);
				\draw (v2221)--(v22212);
				\draw (r)--(v3);
				\draw (v3)--(v31);
				\draw (v31)--(v311);
				\draw (v31)--(v312);
				\draw (v31)--(v313);
				\draw (v3)--(v32);
				
				\node[inner sep = -11pt, outer sep = 0 pt] (c1) at (-2.1,0.95) {\usebox\Xa};
				\node[inner sep = -9pt, outer sep = 0 pt] (c2) at (-0.4,-1.15) {\usebox\Xb};
				\node[inner sep = -5pt, outer sep = 0 pt] (c3) at (3.25,-1.7) {\usebox\Xc};
				\node[inner sep = -5pt, outer sep = 0 pt] (c4) at (2.55,1.65) {\usebox\Xd};
				
				\draw[shorten <= 5pt] (c1) to[out = 30,in = 90] (v12);
				\draw (c1) -- (v12);
				\draw[shorten <= 3pt] (c1) to[out = -90,in = 180] (v12);
				\draw[shorten <= 7pt] (c1) to[out = -100,in = 160] (v11);
				
				\draw[] (c2) to[out = 160,in = -50] (v11);
				\draw[] (c2) to[out = -5,in = -160] (v21);
				
				\draw[] (c3) to[out = 110,in = -100] (v2221);
				\draw[] (c3) to[out = 130,in = -60] (v222);
				
				\draw[] (c4) to[out = -160, in = 50, shorten <= 10pt] (v32);
				\draw[] (c4) to[out = -180,in = 90] (v32);
				\draw[] (c4) to[out = -150,in = 80] (v31);
				\draw[] (c4) to[out = -30,in = 30] (v311);
				\draw[] (c4) to[out = -90,in = 65] (v313);
				\end{tikzpicture}
			}
		}
		\end{minipage}
		\hfill\begin{minipage}{0.5\textwidth}
			\resizebox{\textwidth}{!}{
			\begin{tikzpicture}
			\useasboundingbox (-2.0525,-2.3) rectangle (4.4025,2.3);
			
			\node[vertex] (r) at (0,0) {};
			\node[vertex] (v1) at (-0.8,0) {};
			\node[vertex] (v11) at (-1.55,-0.5) {};
			\node[vertex] (v12) at (-1.55,0.5) {};
			\node[vertex] (v2) at (0.75,-0.5) {};
			\node[vertex] (v21) at (1.1,-1.1) {};
			\node[vertex] (v22) at (1.55,-0.5) {};
			\node[vertex] (v221) at (1.9,-1.1) {};
			\node[vertex] (v222) at (2.35,-0.5) {};
			\node[vertex] (v2221) at (3.15,-0.5) {};
			\node[vertex] (v22211) at (3.85,0) {};
			\node[vertex] (v22212) at (3.85,-1) {};
			\node[vertex] (v3) at (0.75,0.5) {};
			\node[vertex] (v31) at (1.55,0.5) {};
			\node[vertex] (v311) at (2.35,0) {};
			\node[vertex] (v312) at (2.35,0.5) {};
			\node[vertex] (v313) at (2.35,1) {};
			\node[vertex] (v32) at (1.4,1) {};
			\draw (r)--(v1);
			\draw (v1)--(v11);
			\draw (v1)--(v12);
			\draw (r)--(v2);
			\draw (v2)--(v21);
			\draw (v2)--(v22);
			\draw (v22)--(v221);
			\draw (v22)--(v222);
			\draw (v222)--(v2221);
			\draw (v2221)--(v22211);
			\draw (v2221)--(v22212);
			\draw (r)--(v3);
			\draw (v3)--(v31);
			\draw (v31)--(v311);
			\draw (v31)--(v312);
			\draw (v31)--(v313);
			\draw (v3)--(v32);
			
			\node[vertex, red] at (v12) {};
			
			\node[vertex, red] at (v11) {};
			\node[vertex, red] at (v21) {};
			
			\node[vertex, red] at (v222) {};
			\node[vertex, red] at (v2221) {};
			
			\node[vertex, red] at (v31) {};
			\node[vertex, red] at (v32) {};
			\node[vertex, red] at (v311) {};
			\node[vertex, red] at (v313) {};
			
			\draw[red, dashed] (v11)--(v12);
			\draw[red, dashed] (v11) to[out = -50,in = -160] (v21);
			
			\draw[red, dashed] (v31)--(v32);
			\draw[red, dashed] plot[smooth,tension=1] coordinates{(v32) (2.25,1.3) (2.9,.7) (v311)};
			\draw[red, dashed] (v32)--(v313);
			\draw[red, dashed] (v311) to[out = 50,in = -50] (v313);
			\end{tikzpicture}
		}
		\end{minipage}
		\vspace{-0.7cm}
		\caption{
			\emph{Left:} A graph $G$ with a tree as a modulator to $\cH$ (the part in $\cH$ is depicted hatched).
			\emph{Right:} The corresponding $\cH$-torso.}
			\vspace{-0.2cm} 
\label{fig:htw}
	\end{figure}

\begin{definition}[\(\cH\)-Treewidth]
	The \(\cH\)-treewidth of a graph \(G\) is the minimum treewidth of an \(\cH\)-torso of \(G\).
	We denote the \(\cH\)-treewidth of \(G\) by \(\classtreewidth{\cH}(G)\).
\end{definition}

\iflong
Typically, we will want to consider a graph class \(\cH\) for which certain problems are polynomial-time tractable. Hence, we will assume w.l.o.g.\ that  \((\emptyset, \emptyset) \in \cH\).
From the definition it is obvious that:
\begin{observation}
	\label{obs:treewidth-bound}
	For any graph \(G\), \( \classtreewidth{\cH}(G) \leq \tw(G)\).
\end{observation}
\fi
\ifshort
Typically, we will want to consider a graph class \(\cH\) for which certain problems are polynomial-time tractable. Hence, we will assume w.l.o.g.\ that  \((\emptyset, \emptyset) \in \cH\).
From the definition we easily observe that \( \classtreewidth{\cH}(G) \leq \tw(G)\) for every $G$ and $\cH$.
\fi

\subsection{Nice \(\cH\)-Tree-Decompositions}
Just like for tree decompositions, we can also define a canonical form of decompositions which has properties that are convenient when formulating dynamic programs using \(\cH\)-treewidth. Intuitively, a \emph{nice \(\cH\)-tree decomposition} behaves like a nice tree decomposition on the torso graph (see points 1-3), with the exception that the neighborhoods of the collapsed parts must occur as special \emph{boundary} leaves (see points 4-5).

\ifshort
\begin{definition}[Nice \(\cH\)-Tree-Decomposition]
	A nice \(\cH\)-tree decomposition of a graph \(G\) is a triple \((X, T, \{B_t \mid t \in V(T)\})\) where
	\(X \subseteq V(G)\) such that \(G \circ X\) is an \(\cH\)-torso,
	\((T, \{B_t \mid t \in V(T)\})\) is a rooted tree decomposition of \(G \circ X\), and:
	\begin{enumerate}
		\item Every node in \(T\) has at most two children.
		\item If a node \(t\) has children \(t_1 \neq t_2\), then \(B_t = B_{t_1} = B_{t_2}\) and we call \(t\) a \emph{join node}.		
		\item If a node \(t\) has exactly one child \(t'\), then either (a) there exists $x\in V(G)\setminus B_{t'}$ such that $B_t= B_{t'}\cup \{x\}$ and we call \(t\) an \emph{introduce node}, or (b) there exists $x\in V(G) \setminus B_t$ such that $B_{t'}= B_{t}\cup \{x\}$ and we call $t$ a \emph{forget node}.
		\item If a node \(t\) is a leaf, then (a) \(\vert B_t \vert = 1\) and we call \(t\) a \emph{simple leaf}, or (b) \(B_t = N(C)\) for some connected component \(C\) of \(G[X]\) and we call \(t\) a \emph{boundary leaf}.
		\item For each connected component \(C\) of \(G[X]\) there is a unique leaf \(t\) with \(B_{t} = N(C)\).		
	\end{enumerate}
\end{definition}
\fi

\iflong
\begin{definition}[Nice \(\cH\)-Tree-Decomposition]
	A nice \(\cH\)-tree decomposition of a graph \(G\) is a triple \((X, T, \{B_t \mid t \in V(T)\})\) where
	\(X \subseteq V(G)\) such that \(G \circ X\) is an \(\cH\)-torso,
	\((T, \{B_t \mid t \in V(T)\})\) is a rooted tree decomposition of \(G \circ X\), and:
	\begin{enumerate}
		\item Every node in \(T\) has at most two children.
		\item If a node \(t\) of \(T\) has two children \(t_1\) and \(t_2\), then \(B_t = B_{t_1} = B_{t_2}\);
		in this case we call \(t\) a \emph{join node}.		
		\item If a node \(t\) of \(T\) has exactly one child \(t'\), then either of the following holds:
		\begin{enumerate}
			\item \(\vert B_t \vert = \vert B_{t'} \vert + 1\), in which case we call \(t\) an \emph{introduce node}, or
			\item \(\vert B_t \vert = \vert B_{t'} \vert - 1\) in which case we call \(t\) a \emph{forget node}.
		\end{enumerate}		
		\item If a node \(t\) is a leaf, then one of the following holds:
		\begin{enumerate}
			\item \(\vert B_t \vert = 1\); in which case we call \(t\) a \emph{simple leaf node}, or
			\item \(B_t = N(C)\) for some connected component \(C\) of \(G[X]\); in which case we call \(t\) a \emph{boundary leaf node}.
		\end{enumerate}
		\item For each connected component \(C\) of \(G[X]\) there is a unique leaf \(t\) with \(B_{t} = N(C)\).		
	\end{enumerate}
\end{definition}
\fi

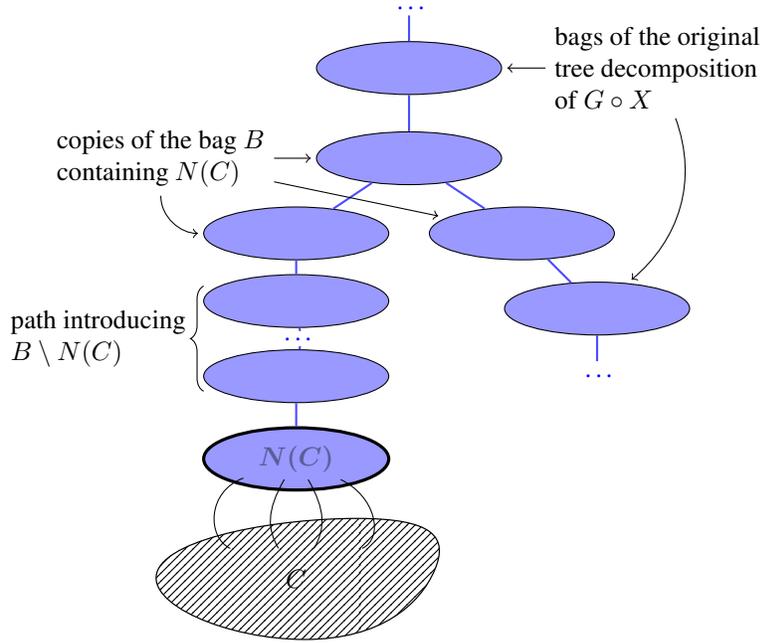
\begin{figure}[ht]
	
		\centering
		\begin{tikzpicture}
		\useasboundingbox (-3.6,-7.4) rectangle (3.6,0.8);
		
			\node at (0.05,0.8) {\color{blue} \dots};
			\node[bag] (B1) at (0,0) {};
			\node[bag] (B3) at (2.5,-3.2) {};
			\node at (2.55, -4.1) {\color{blue} \dots};
			\node[bag] (B2) at (1.5,-2.2) {};
			\node[bag] (B2j) at (0,-1.2) {};
			\node[bag] (B2') at (-1.5,-2.2) {};
			\node[bag] (F1) at (-1.5,-3.1) {};
			\node at (-1.45, -3.6) {\color{blue} \dots};
			\node[bag] (FB-N) at (-1.5,-4.1) {};
			\node[bag, very thick] (boundary) at (-1.5,-5.2) {\(\bm{N(C)}\)};
			\draw[thick, blue!70] (0,0.7)--(B1);
			\draw[thick, blue!70] (B3)--(2.5, -3.9);
			
			\draw[thick, blue!70] (B1)--(B2j);
			\draw[thick, blue!70] (B2j)--(B2);
			\draw[thick, blue!70] (B2j)--(B2');
			\draw[thick, blue!70] (B2)--(B3);
			\draw[thick, blue!70] (B2')--(F1);
			\draw[thick, blue!70] (F1)--(-1.45, -3.5);
			\draw[thick, blue!70] (-1.45, -3.7)--(FB-N);
			\draw[thick, blue!70] (FB-N)--(boundary);
			
			\node[inner sep = -5pt, outer sep = 0 pt, scale = 2.3] (C) at (-1.5,-6.8) {\usebox\C};
			\node at (-1.5,-6.8) {\(C\)};
			\draw[shorten <= -5pt] (boundary) to[out = -25, in = 25] (C);
			\draw[shorten <= -5pt] (boundary) to[out = -60, in = 60] (C);
			\draw[shorten <= -5pt] (boundary) to[out = -120, in = 120] (C);
			\draw[shorten <= -5pt] (boundary) to[out = -160, in = 155] (C);
			
		 	\node[align=left] (original-bags) at (3.3,0)
		 	{
		 		bags of the original\\
		 		tree decomposition\\
		 		of \(G \circ X\)
	 		};
 			\draw[->, shorten >= 2pt] (original-bags) to (B1);
 			\draw[->, shorten >= 2pt] (original-bags) to[out = -70, in = 40] (B3);
 			\node[align=left] (copy-bags) at (-3.3,-1.2)
 			{
 				copies of the bag $B$\\
 				containing \(N(C)\)
 			};
 			\draw[->, shorten >= 2pt] (copy-bags) to (B2j);
 			\draw[->, shorten >= 3pt] (copy-bags) to (B2);
 			\draw[->, shorten >= 2pt] (copy-bags) to[out = -90, in = 180] (B2');
 			\draw[decorate,decoration={brace,amplitude=5pt},xshift=-35pt,yshift=0pt]
 			(-1.5,-4.3) -- (-1.5,-2.9) node [align=left, black,midway,xshift=-40pt] 
 			{
 				path introducing\\
 				\(B \setminus N(C)\)
 			};

		\end{tikzpicture}
		\caption{
			Part of a nice \(\cH\)-tree-decomposition (blue) including a boundary leaf (bold) and a connected component \(C\) (hatched) of $X$.
		}\label{fig:nicedec}
\end{figure}
An illustration of a nice \(\cH\)-tree decomposition showcasing how it differs from a nice tree decomposition is provided in Figure~\ref{fig:nicedec}; in line with standard terminology for treewidth, we call the sets $B_t$ \emph{bags}.
The width of a nice \(\cH\)-tree decomposition is simply the width of \((T, \{B_t \mid t \in V(T)\})\).
\ifshort
Given a node $t$ in a nice $\cH$-tree decomposition $T$, we let $Y_t$ be the set of all vertices contained in the bags of the subtree rooted at $t$, i.e., $Y_t=B_t\cup \bigcup_{p \text{ is separated from the root by $t$}}B_p$.
It is possible to show that computing a nice \(\cH\)-tree decomposition of bounded width can be reduced to finding an appropriate \(\cH\)-torso (this is because a nice $\cH$-tree decomposition can be obtained straightforwardly from a nice tree decomposition of the torso). $(\clubsuit)$
\fi
\iflong
Below we show that computing a nice \(\cH\)-tree decomposition of bounded width can be reduced to finding an appropriate \(\cH\)-torso.

\begin{lemma}
	Given an $n$-vertex graph \(G\) and an \(\cH\)-torso \(U\) of \(G\) with \(\tw(U) \leq k\), we can find a nice \(\cH\)-tree decomposition of \(G\) with width at most \(k\) in time $k^{\bigoh(k^3)}\cdot n$.
\end{lemma}
\begin{proof}
	Compute a nice tree decomposition \((T, \{B_t \mid t \in V(T)\})\) of \(U\) with width at most \(k\) in time at most $k^{\bigoh(k^3)}\cdot n$ using Fact~\ref{fact:findtw}.
	Let \(X := V(G) \setminus V(U)\).
	By the definition of \(\circ\), \(U = G \circ X\) and
	the neighbors of each connected component \(C\) of \(G[X]\) form a clique in \(U\).
	Thus, for each \(C\) there is some node, say \(t_C\), of \(T\) such that \(N(C) \subseteq B_t\).
	
	Obtain \(T'\) from \(T\) by successively performing the following, for each connected component \(C\) of \(G[X]\):
	Replace \(t_C\) by a new node \(t'\) with two children \(t_1\) and \(t_2\),
	and set \(B_{t'} = B_{t_1} = B_{t_2} = B_{t_C}\).
	Below \(t_1\), attach the descendants of \(t_C\) (i.e., a copy of the subtree rooted at $t_C$ taken from the tree that was obtained after the adaptations for the previously considered connected components of \(G[X]\)),
	and below \(t_2\) attach a path \(s_1, \dotsc, s_{\vert B_t \setminus N(C) \vert}\).
	Set \(B_{s_1} = B_{t_2} \setminus \{v_1\}\) and \(B_{s_{i+1}} = B_{s_i} \setminus \{v_{i+1}\}\) where
	\(\{v_1, \dotsc, v_{\vert B_t \setminus N(C) \vert}\}\) is some enumeration of \(B_t \setminus N(C)\).
	
	It is easy to see that these manipulations result in a nice \(\cH\)-tree decomposition without introducing bags larger than the ones already occuring in \((T, \{B_t \mid t \in V(T)\})\).
\end{proof}
\fi

Hence, we can state the problem of computing a decomposition as follows:
\begin{center}
	\begin{mathproblem}{$\cH$-Treewidth}[\(k\)]
		Instance: & A graph $G$, an integer $k$.\\
		Task: & Find an $\cH$-torso $U$ of $G$ such that $\tw(U) \leq k$, or correctly determine that no such $\cH$-torso exists.\\
	\end{mathproblem}
	\vspace{-0.3cm}
\end{center}

\iflong \subsection{An Algorithmic Meta-Theorem}\fi

\ifshort \noindent \textbf{An Algorithmic Meta-Theorem. }\fi
Before proceeding to the flagship application of $\cH$-treewidth where $\cH$ is the class of graphs of bounded rank-width, here
we give a generic set of conditions that allow fixed-parameter algorithms for problems parameterized by \(\cH\)-treewidth.
Specifically, we consider graph problems that are \emph{finite-state}~\cite{Courcelle90}
 or have \emph{finite integer index}~\cite{BodlaenderFominLokshtanovPenninkxSaurabhThilikos16,BodlaenderF01,GajarskyHOORRVS17}. Informally speaking, such problems only transfer a limited amount of information across a small separator in the input graph and hence can be solved ``independently'' on both sides of such a separator. Since these notions are only used in this section, we provide concise definitions below.

First of all, we will need the notion of boundaried graphs and gluing. A graph $\bar G$ is called \emph{\(t\)-boundaried} if it contains $t$ distinguished vertices identified as $b^G_1,\dots, b^G_t$. The \emph{gluing operation} $\oplus$ takes two $t$-boundaried graphs $\bar G$ and $\bar H$, creates their disjoint union, and then alters this disjoint union by identifying the boundaries of the two graphs (i.e.\ by setting $b^G_i=b^H_i$ for each $i\in [t]$).

Consider a decision problem $\pP$ whose input is a graph. We say that two $t$-boundaried graphs $\bar C$ and $\bar D$ are \emph{equivalent}, denoted by $\bar C \sim_{\pP,t} \bar D$, if for each $t$-boundaried graph $\bar H$ it holds that 
\ifshort $\bar C \oplus \bar H\in \pP \text{ if and only if } \bar D \oplus \bar H\in \pP.$ \fi
\iflong
\[
	\bar C \oplus \bar H\in \pP \text{ if and only if } \bar D \oplus \bar H\in \pP.
\]
\fi
We say that $\pP$ is \emph{finite-state} (or \emph{FS}, in brief) if, for each $t\in \Nat$, $\sim_{\pP,t}$ has a finite number of equivalence classes.\\
Next, consider a decision problem $\pQ$ whose input is a graph and an integer. In this case we say that two $t$-boundaried graphs $\bar C$ and $\bar D$ are equivalent (denoted by $\bar C \sim_{\pQ,t} \bar D$) if there exists an \emph{offset} $\delta(\bar C, \bar D)\in \Int$ such that for each $t$-boundaried graph $\bar H$ and each $q\in \Int$:
\iflong
\begin{center}
	$(\bar C \oplus \bar H, q)\in \pQ$ if and only if $(\bar D \oplus \bar H,q+\delta(\bar C, \bar D))\in \pQ.$
\end{center}
\fi
\ifshort 	$(\bar C \oplus \bar H, q)\in \pQ$ if and only if $(\bar D \oplus \bar H,q+\delta(\bar C, \bar D))\in \pQ.$
\fi
We say that $\pQ$ has \emph{finite integer index} (or is \emph{FII}, in brief) if, for each $t\in \Nat$, $\sim_{\pQ,t}$ has a bounded number of equivalence classes.

We note that a great number of natural graph problems are known to be FS or FII. For instance, all problems definable in Monadic Second Order logic are FS~\cite[Lemma 3.2]{BodlaenderFominLokshtanovPenninkxSaurabhThilikos16},
while examples of FII problems include \textsc{Vertex Cover}, \textsc{Independent Set}, \textsc{Feedback Vertex Set}, \textsc{Dominating Set}, 
\iflong
\textsc{Connected Dominating Set}, and \textsc{Edge Dominating Set}~\cite{GajarskyHOORRVS17}.

\fi
\ifshort to name a few~\cite{GajarskyHOORRVS17}.\fi
We say that a FS or FII problem $\pP$ is \emph{efficiently extendable} on a graph class $\cH$ if there is a fixed-parameter algorithm (parameterized by $t$) that takes as input a $t$-boundaried graph $\bar G$ such that the boundary is a modulator to $\cH$ and outputs the equivalence class of $\bar G$ w.r.t.\ $\sim_{\pP,t}$.

\ifshort
\begin{theorem}[$\clubsuit$]
	\label{thm:meta}
	Let $\pP$ be a FS or FII graph problem and $\cH$ be a graph class such that (1) $\pP$ is efficiently extendable on $\cH$, (2) $\pP$ is \FPT parameterized by treewidth, and
(3) \textsc{$\cH$-treewidth} is \FPT.
	Then $\pP$ is \FPT parameterized by $\cH$-treewidth.
\end{theorem}

\begin{proof}[Proof Sketch]
We can solve \(\pP\) as follows.
	First of all, we use Point (3) to compute an $\cH$-torso $G\circ X$ of treewidth $k$, where $k$ is the $\cH$-treewidth. Next, for each connected component $C$ of $G[X]$, we use the fact that $C\in \cH$ and Point (1) to compute the equivalence class of the boundaried graph $\bar H=\overline{G[C\cup N(C)]}$ where the boundary is $N(C)$. Note that since $\tw(G\circ X)\leq k$ and $N(C)$ forms a clique in $G\circ X$, $|N(C)|\leq k$ and hence this step takes only fixed-parameter time.	
	Next, we use a brute-force enumeration argument to compute a bounded-size representative of the equivalence class of $\bar H$, and replace $\bar H$ with this representative. After doing this exhaustively, we obtain a graph $G'$ of bounded treewidth, for which we can invoke Point (2).
\end{proof}
\fi
\iflong\begin{theorem}
	\label{thm:meta}
	Let $\pP$ be a FS or FII graph problem and $\cH$ be a graph class. If
	\begin{enumerate}
		\item $\pP$ is efficiently extendable on $\cH$,
		\item $\pP$ is \FPT parameterized by treewidth, and
		\item \textsc{$\cH$-treewidth} is \FPT,
	\end{enumerate}
	then $\pP$ is \FPT parameterized by $\cH$-treewidth.
\end{theorem}

\begin{proof}
	\(\pP\) can be solved using the following algorithm.
	First of all, we use Point~3. to compute an $\cH$-torso $G\circ X$ of treewidth $k$, where $k$ is the $\cH$-treewidth of the input graph $G$. Next, for each connected component $C$ of $G[X]$, we use the fact that $C\in \cH$ and Point 1. to compute the equivalence class of the boundaried graph $\bar H=\overline{G[C\cup N(C)]}$ where the boundary is $N(C)$. Note that since $\tw(G\circ X)\leq k$ and $N(C)$ forms a clique in $G\circ X$, $|N(C)|\leq k$ and hence this step takes only fixed-parameter time.
	
	Our next task is to compute a minimum-size representative of the equivalence class $\alpha$ of $\bar H$; we remark that the procedure used to do so in the general setting of this meta-theorem does not allow us to give any explicit upper bound on the parameter dependency. We describe the procedure for the case of finite-state problems; FII problems are handled analogously, with the sole distinction being that we also keep track of the offset $\delta$. In particular, we use a brute-force enumeration argument that has previously been applied in the kernelization setting~\cite{GanianSS16}: we enumerate all boundaried graphs (by brute force and in any order with a non-decreasing number of vertices), and for each graph we check whether its equivalence class w.r.t.\ $\sim_{\pP,k}$ is $\alpha$. Observe that since the size of a minimum-cardinality representative of $\alpha$ depends only on $\pP$ and $k$, the number of graphs that need to be checked in this way as well as the size of each such graph depends only on $\pP$ and $k$, and hence there exists some function of $\pP$ and $k$ which upper-bounds the runtime of this brute-force procedure. Let $\bar G_\alpha$ be the computed minimum-cardinality representative of $\alpha$.
	
	In the final step, for each $C$ the algorithm glues $\overline{G\circ X}$ (with boundary $N(C)$) with the computed representative $\bar G_\alpha$. Let $G'$ be the graph obtained after processing each connected component $C$ in the above manner. Note that since $\tw(G\circ X)\leq k$ and each $\bar G_\alpha$ has size bounded by a function of $k$, the graph $G'$ has treewidth bounded by a function of $k$. Indeed, one can extend a tree decomposition of $G\circ X$ by adding each $\bar G_\alpha$ into a new bag (a leaf in the decomposition) adjacent to the leaf of the nice \(\cH\)-tree decomposition we obtain containing $N(C)$. At this stage, the algorithm invokes Point 2. and outputs \yes\ iff $G'\in \pP$. In the case of FII problems where the input was $(G,\ell)$, the algorithm instead outputs \yes\ iff $(G',\ell')\in \pP$, where $\ell'$ is obtained from $\ell$ by subtracting all the individual offsets $\delta(G[C\cup N(C)],G_\alpha)$. Correctness follows by the fact that $G_\alpha \sim_{\pP,k} G[C\cup N(C)]$.
\end{proof}
\fi

\section{\(\cRc{c}\)-Treewidth}
This section focuses on the properties of \(\cRc{c}\)-treewidth, a hierarchy of graph parameters that represent our flagship application of the generic notion of $\cH$-treewidth.

\iflong \subsection{Comparison to Known Parameters} 
It follows from Observation~\ref{obs:treewidth-bound} that $\cRc{c}$-treewidth dominates treewidth (for every $c\in \Nat$).  \fi
\ifshort 
\smallskip
\noindent \textbf{Comparison to Known Parameters.} It follows from the definition of $\cH$-treewidth that $\cRc{c}$-treewidth dominates treewidth (for every $c\in \Nat$).
\fi
Similarly, it is obvious that $\cRc{c}$-treewidth dominates the size of a modulator to $\cRc{c}$ (also for every $c\in \Nat$).
The following lemma shows that, for every fixed \(c\), \(\cRc{c}\)-treewidth is dominated by rankwidth.
\begin{lemma}\
\label{lem:belowrw}
Let \(c\in \Nat\). If $\classtreewidth{\cRc{c}}(G)=k$ then \(\rw(G) \leq c + k + 1\).
\end{lemma}
\begin{proof}
	Let the \(\cRc{c}\)-treewidth of \(G\) be witnessed by some nice \(\cRc{c}\)-tree-decomposition \((X, T, \{B_t \mid t \in V(T)\})\) of width \(k\).
	
	We can obtain a rank-decomposition \((T', \mu)\) of \(G\) from \((X, T, \{B_t \mid t \in V(T))\}\) as follows:
	
	For vertices \(v\) of \(G \circ X\) such that there is no leaf node \(t \in V(T)\) with \(B_t = \{v\}\), let \(t \in V(T)\) be a forget node with child \(t'\) such that \(B_t \cup \{v\} = B_{t'}\).
	Turn \(t'\) into a join node by introducing \(t_1\), \(t_2\) with \(B_{t_1} = B_{t_2} = B_{t'}\) as children of \(t'\), attaching the former child of \(t'\) to \(t_1\) and a new leaf node \(t_v\) with \(B_{t_v} = \{v\}\) below \(t_2\).
	Note that this preserves the fact that for any \(v \in V(G) \setminus X\), \(T[\{u \in V(T) \mid v \in B_u\}]\) is a tree.
	Now we can choose for each \(v \in V(G \circ X)\) some \(\mu(v) \in V(T)\) such that \(B_t = \{v\}\).
	This defines an injection from \(V(G \circ X)\) to the leaves of \(T\).
	However not every leaf of \(T\) is mapped to by \(\mu\).
	On one hand there are the boundary leaf nodes, below which we will attach subtrees to obtain a rank-decomposition of \(G\).
	On the other hand there may be \(v \in V(G \circ X)\) for which the choice of \(\mu(v)\) was not unique, i.e.\ there is \(t \neq \mu(v)\) with \(B_t = \{v\}\).
	For all such \(v\) and \(t\) we delete all nodes on the root-\(t\)-path in \(T\) that do not lie on a path from the root to a vertex in \(\{\mu(w) \mid w \in V(G \circ X)\} \cup \{t' \mid t' \text{ boundary leaf node}\}\).
	This turns \(\mu\) into an injection from \(V(G \circ X)\) to the leaves of \(T\), that is surjective on the non-boundary leaf nodes.
	
	Next, we extend \((T, \mu)\) to a rank-decomposition of \(G\) by proceeding in the following way for each connected component \(C\) of \(G[X]\):\\
	Let \(t_C \in V(T)\) be the boundary leaf node with \(B_t = N(C)\).
	Since \(\rw(C) \leq c\), we find a rank-decomposition \((T_C, \mu_C)\) of \(C\) with width at most \(c\).
	Attach \(T_C\) below \(t_C\).\\
	Let \(T'\) be the tree obtained by performing these modifications for all connected components \(C\).
	Consider the rank-decomposition of \(G\) given by
	\[\left(T', v \mapsto
	\begin{cases}
	\mu(v) & \mbox{if } v \in V(G \circ X)\\
	\mu_C(v) & \mbox{if } 
	\begin{aligned}[t]
	& v \in C\text{ for some }C \text{ as above}
	\end{aligned}
	\end{cases}\right).\]
	 
	We show that its width is at most \(c + k + 1\).
	Any edge \(e\) of \(T'\) is of one of the following types:
	\begin{itemize}
		\item \(e\) corresponds to an edge already contained in \(T\):
		Then \(e\) induces a bipartition \((X_G, Y_G)\) of \(V(G)\).
		Fix \(t \in V(T)\) to be the vertex in which \(e\) starts.\\
		Let \(x \in X_G\) and \(y \in Y_G\) be such that \(xy \in E(G)\).
		Observe that \(e\) does not separate neighbors in \(X\) as these lie within the same connected component of \(G[X]\) whose rank-decomposition is, by construction, attached completely within one of the two subtrees of \(T'\) separated by \(e\).
		So, we consider \(x \in V(G \circ X)\).
		If \(y \in V(G \circ X)\) then \(x\) and \(y\) occur together in some bag of the original tree of the tree decomposition, and as remarked earlier this is still the case modified tree.
		This implies that at least one of \(\{x,y\}\) must be present in \(B_t\).
		If \(y \notin V(G \circ X)\), by the construction of \(T'\), \(y\) corresponds to a leaf \(t_y \in V(T')\) in a subtree attached to \(T\) rooted at \(t_C \in V(T)\) with \(B_{t_C} = N(C) \ni x\) and \(t_C\) and \(t_y\) are not separated by the removal of \(e\).
		Also, \(x\) corresponds to a leaf \(t_x \in V(T)\) which is in the subtree of \(T' - e\) not containing \(t_y\), i.e.\ the subtree not containing \(t_C\).
		This means any subtree containing \(t_C\) and \(t_x\) also contains \(t\), and
		since \((T, \{B_t \mid t \in V(T)\})\) is a tree-decomposition and \(x \in B_{t_x} \cap B_{t_C}\) this means \(x \in B_t\).\\
		In both cases at least one of \(\{x,y\}\) is in \(B_t\).
		Since this argument applies for every edge crossing the bipartition $(X_G,Y_G)$, it follows that $\bm{A}_G[X_G,Y_G]$ may only contain ``1'' entries in rows and columns that correspond to the vertices in $B_t$. Since $|B_t|\leq k+1$, it holds that $\bm{A}_G[X_G,Y_G]$ can be converted into a zero matrix by deleting at most $k+1$ rows plus columns, which is a sufficient condition for $\bm{A}_G[X_G,Y_G]$ having rank at most $k+1$.
		\item \(e\) corresponds to an edge in a rank-decomposition of some connected component \(C\) of \(G[X]\):
		Then \(e\) induces a bipartition \((X_G, Y_G)\) of \(V(G)\) and a bipartition \((X_C, Y_C)\), where \(X_C = X_G \cap C\) and \(Y_C = Y_G \cap C\), of \(C\).
		Since vertices of \(C\) are only connected to vertices in \(N(C)\) outside of \(C\) and \(N(C) \subseteq B_t\) for some \(t \in V(T)\), we have
		\(\rho_G(X_G) \leq \rho_C(X_C) + \vert N(C) \vert \leq c + k + 1\).
		\item \(e\) corresponds to an edge connecting the rank-decomposition of some connected component \(C\) of \(G[X]\) to \(T\):
		Then the bipartition induced is \((C, V(G) \setminus C)\) and as \(N(C) \subseteq B_t\) for some \(t \in V(T)\) \(\rho_G(C) \leq \vert N(C) \vert \leq k + 1\).\qedhere
	\end{itemize}
\end{proof}

\ifshort
	Next, we compare \(\cRc{c}\)-treewidth to Telle and Saether's \emph{sm-width}~\cite{SaetherTelle14} ($\clubsuit$).
\begin{lemma}[$\clubsuit$]
		\(\cRc{c}\)-treewidth and sm-width are incomparable.
	\end{lemma}
	
\fi

	\iflong
	Next, we compare \(\cRc{c}\)-treewidth to Telle and Saether's notion of \emph{sm}-width~\cite{SaetherTelle14}, which is another parameter that lies between treewidth and rank-width. 
	The rough idea behind the definition of this parameter is to generalize treewidth towards denser graph classes by allowing the use of \emph{graph splits} in an explicit way without increasing the parameter value. For completeness, we provide a formal definition for sm-width below using the general notions of \emph{branch-width} and graph splits.

	\begin{definition}[Branch-Width~\cite{RobertsonSeymour91}]
		A branch decomposition of the vertex set of a graph \(G\) is a binary tree \(T\) with \(V(G)\) as leaf set. Each edge $e$ in $T$ partitions the vertices of $V(G)$ into two sets (say $X_e$ and $V(G)\setminus X_e$), corresponding to the leaves present in the two connected components of $T-e$.
		
		A cut function is a function \(c : \mathcal{P}(V(G)) \to \Nat\) such that
		\[
		\forall A \subseteq V(G) \ c(A) = c(V(G) \setminus A).
		\]		
		The width of a branch decomposition with respect to the cut function \(c\) is given by
		$
		\max_{e \in E(T)}c(X_e).
		$		
		The branch width of a graph with respect to a cut function is the minimum of the branch width with respect to the cut function over all branch decompositions of the vertex set of this graph.
	\end{definition}
	
	\begin{definition}[Split]
		\(A \subseteq V(G)\) is a split of a graph \(G\) if
		\begin{enumerate}
			\item \(|A| \geq 2\), \(|V(G) \setminus A| \geq 2\) and
			\item
			\(\begin{aligned}[t]
			\forall v, w \in A \ N(v) \cap (V(G) \setminus A) \neq \emptyset \land N(w) \cap (V(G) \setminus A) \neq \emptyset \\
			\Rightarrow N(v) \cap (V(G) \setminus A) = N(w) \cap (V(G) \setminus A).
			\end{aligned}\)
		\end{enumerate}
	\end{definition}
	
	\begin{definition}[Split-Matching-Width~\cite{SaetherTelle14}]
		The split-matching-width (or sm-width) of a graph \(G\), denoted by \(\operatorname{smw}(G)\), is the branch width of \(G\) with respect to the cut function given by
		\[
		c(A) =
		\begin{cases}
		1 & \mbox{if \(A\) is a split}\\
		\max\{\vert M \vert \mid M \text{is a matching of } G[A, V(G) \setminus A]\} & \mbox{otherwise}
		\end{cases}.
		\] 
	\end{definition}
	
	For clarity, we remark that omitting the first case for $c(A)$ in the definition of sm-width results in a parameter called \emph{mm-width} which is asymptomatically equivalent to treewidth~\cite{JeongST18}.
	
\begin{lemma}
		\(\cRc{c}\)-treewidth and \(\operatorname{smw}\) are incomparable.
		More formally, the following hold:
		\begin{enumerate}
			\item \label{tw_below_smw} 
			For all \(c \in \Nat\) there is a sequence of graphs \((G_i)_{i \in \Nat}\) such that
			for all \(i \in \Nat\) \(\classtreewidth{\cRc{c}}(G_i) \geq i\),
			but  \(\operatorname{smw}(G_i) \leq c + 3\).
			\item \label{smw_below_tw} 
			For all \(c\geq 3\) there is a sequence of graphs \((G_i)_{i \in \Nat}\) such that
			for all \(i \in \Nat\)
			\(\operatorname{smw}(G_i) \geq i\), but 
			\(\classtreewidth{\cRc{c}}(G_i) = 0\).
		\end{enumerate}
	\end{lemma}
	\begin{proof}
		For \ref{tw_below_smw} consider the following construction:
		For arbitrary \(c\) there is a graph \(H_c\) such that $\tw(H_c) = c+2$ and $\rw(H_c) = c + 1$; notably, this can be achieved by setting $H_c$ to be the $(c + 2)\times (c + 2)$ square grid~\cite{Jelinek10}.
		Now let \(G_i\) be given by taking \(2 \cdot (i + 1)\) copies, \(H_c^{(1)}, \dotsc, H_c^{(2 \cdot (i + 1))}\) of \(H_c\) and completely connecting all \(H_c^ {(j)}\) with \(j\) even to all \(H_c^{(j)}\) with \(j\) odd.\\
		The split-matching-width of \(G_i\) is at most \(c + 3\) since each copy of \(H_c\) iteratively defines a split and thus can be treated independently and it is known that the maximum matching width of a graph is at most its treewidth\(\,+ 1\)~\cite{Vatshelle12}.\\
		On the other hand, we claim that \(\classtreewidth{\cRc{c}}(G_i) \geq i\):
		\begin{itemize}
			\item 
			If \(G_i\) has no non-trivial \(\cRc{c}\)-torso,
			then since \(G_i\) is at least \(((i + 1) \cdot |V(H_c)|)\)-degenerate because of the complete connections between odd and even copies of \(H_c\),
			we know that
			\[
			\classtreewidth{\cRc{c}}(G_i)
			= \tw(G_i)
			\geq (i + 1) \cdot |V(H_c)|
			\geq i.
			\]
			\item
			If the \(\cRc{c}\)-treewidth of \(G_i\) is given by \(\tw(G_i \circ X)\) and it hols that \(X\) only intersects odd (symmetrically only even) copies of \(H_c\) then,
			by definition of \(G_i \circ X\),
			the even (respectively odd) copies of \(H_c\) form a clique in \(G_i \circ X\) and thus
			\[
			\classtreewidth{\cRc{c}}(G_i)
				= \tw(G_i \circ X)
				\geq (i + 1) \cdot |V(H_c)| - 1
				\geq i.
			\]
			\item
			If the \(\cRc{c}\)-treewidth of \(G_i\) is given by \(\tw(G_i \circ X)\) and it holds that
			\(X\) intersects even and odd copies of \(H_c\) and \(|G - X| \geq i + 1\),
			then \(G_i \circ X\) is a clique of size at least \(i + 1\) and thus
			\[
			\classtreewidth{\cRc{c}}(G_i)
				= \tw(G_i \circ X)
				= |V(G_i) \setminus X| - 1
				= i.
			\]
			\item
			Otherwise the \(\cRc{c}\)-treewidth of \(G_i\) is given by \(\tw(G_i \circ X)\) and it holds that \(|G - X| \leq i\).
			Then by the pigeon hole principle we find a \(1 \leq j \leq i\) such that
			\[
			\left\vert V(H_c^{(j)} \cap X) \right\vert
				\geq \frac{|V(G_i)| - i}{2 \cdot (i + 1)}
				= \frac{2 \cdot (i + 1) \cdot |V(H_c)| - i}{2 \cdot (i + 1)}
				> |V(H_c)| - 1.
			\]
			Since \(\rw\left(H_c^{(j)}\right) = c + 1\), this is a contradiction to \(\rw(G[X]) \leq c\).
		\end{itemize}
	
		For \ref{smw_below_tw}, let \(c \geq 3\) and consider the following construction:
		Fix a cycle \(C_i\) on \(3 \cdot i + 5\) vertices.
		Now let \(G_i\) to be the complement graph of \(C_i\).
		Since \(C_i\) has rank-width \(2\) and the complement graph's rank-width is at most one larger~\cite{HlinenyOumSeeseGottlob08}, we know \(\rw(G_i) \leq 3\) and we can use \(V(G_i)\) to build an \(\cRc{c}\)-torso of treewidth \(0\).
		On the other hand, there is no split of \(G_i\)
		and the maximum matching width of a graph \(G\) is known to be at least \(\frac{1}{3} \cdot (\tw(G) + 1)\)~\cite{Vatshelle12}.
		So \(\operatorname{smw}(G_i) \geq \frac{1}{3} \cdot (\tw(G_i) + 1) \geq i\).
	\end{proof}
	\fi

\iflong \subsection{Computing $\cRc{c}$-Treewidth}\fi
\ifshort \smallskip \noindent \textbf{Computing $\cRc{c}$-Treewidth. }\fi
Our aim here is to determine the complexity of computing our parameters, i.e., finding a torso of small treewidth. Obtaining such a torso is a base prerequisite for our algorithms. We formalize the problem below.

\begin{center}
	\begin{mathproblem}{\(\cRc{c}\)-Treewidth}[\(k\)]
		Instance: & A graph \(G\), an integer \(k\).\\
		Task: & Find a \(\cRc{c}\)-torso \(U\) of \(G\) such that \(\tw(U) \leq k\), or correctly determine that no such \(\cRc{c}\)-torso exists.\\
	\end{mathproblem}
\end{center}

\begin{lemma}
\label{lem:computing}
\textsc{$\cRc{c}$-Treewidth} is \FPT.
\end{lemma}

\begin{proof}
We begin by noting that there exists an \MSO\ formula $\phi_c(G)$ that is true iff a graph \(G\) has rank-width at most $c$ (and, equivalently, if each connected component of $G$ has rank-width at most $c$). This was observed already by Kant\'e, Kim, Kwon and Paul~\cite{KanteKKP17}
and follows from the fact that the property of having rank-width at most $c$ can be characterized by a finite set of \emph{vertex minors}~\cite{Oum05},
a property which can be expressed in \MSO~\cite{CourcelleOum07}. For clarity, we explicitly remark that $|\phi_c(G)|$ depends only on $c$.

Next, we also note that there exists an \MSO\ formula $\psi'_k$ (of size depending only on $k$) that is true iff a graph $G$ has treewidth at most $k$~\cite{Lapoire98}.
Consider now the formula $\psi_k(Z)$ for some fixed \(Z \subseteq V(G)\) which operates precisely like $\psi'_k$, but only on the graph $G\circ Z$. In particular, $\psi_k(Z)$ can be obtained from $\psi'_k$ by:
\begin{itemize}
\item For each vertex variable $a$ occurring in $\psi'_k$, adding a condition forcing $v\not \in Z$;
\item For each set variable $A$ occurring in $\psi'_k$, adding a condition forcing $A\cap Z=\emptyset$;
\item Replacing each occurrence of $\text{adj}(a,b)$ with an expression that is true if $\text{adj}(a,b)$ or if there exists an $a$-$b$ path with all internal vertices in $G[Z]$.
\end{itemize}

Let $\phi_{c,k}(Z)=\phi_c(G[Z]) \land \psi_k(Z)$. Consider now a vertex set $Z$ (or, more precisely, an interpretation of $Z$) which satisfies $\phi_{c,k}$. Since $G[Z]$ satisfies $\phi_c$, it holds that each connected component of $G[Z]$ has rank-width at most $c$, and hence $G\circ Z$ is an $\cRc{c}$-torso of $G$. Moreover, since $\psi_k(Z)$ is true, $G\circ Z$ has treewidth at most $k$. 

To conclude the proof, note that Fact~\ref{fact:mso} alongside Lemma~\ref{lem:belowrw} provide an \FPT algorithm (with parameter $k$) to find an interpretation of $Z$ satisfying $\phi_{c,k}(Z)$, or correctly identify that no such interpretation exists.
\end{proof}

\section{Algorithms Exploiting Modulators to $\cRc{c}$}\label{sec:modulator_algorithms}
For a problem $\pP$ to be \FPT\ parameterized by $\cRc{c}$-treewidth, $\pP$ must necessarily be \FPT\ parameterized by treewidth and also \FPT\ parameterized by the size of a modulator to $\cRc{c}$.
However, it is important to note that the latter condition is not sufficient; indeed, one can easily invent artificial problems that are defined in a way which make them trivial in both of the cases outlined above, but become intractable (or even undecidable) once parameterized by $\cRc{c}$-treewidth.
That is, after all, why we need the notion of \emph{efficient extendability} in Theorem~\ref{thm:meta}. 

Hence, in order to develop fixed-parameter algorithms for \textsc{Chromatic Number}, \textsc{Hamiltonian Cycle} and \textsc{Max-Cut} parameterized by $\cRc{c}$-treewidth, we first need to show that they are not only \FPT\ parameterized by the size of a modulator to $\cRc{c}$, but they are also efficiently extendable. Such a result would be sufficient to employ Theorem~\ref{thm:meta} together with Lemma~\ref{lem:computing} in order to establish the desired fixed-parameter tractability results. That is also our general aim in this section, with one caveat: in order to give explicit and tight upper bounds on the parameter dependency of our algorithms, we provide algorithms that solve generalizations of \textsc{Chromatic Number}, \textsc{Hamiltonian Cycle} and \textsc{Max-Cut} parameterized by the size of a modulator to $\cRc{c}$, whereas it will become apparent in the next section that these generalizations precisely correspond to the records required by the treewidth-based dynamic program that will be used in the torso. In other words, the efficient extendability of our problems on $\cRc{c}$ is not proved directly but rather follows as an immediate consequence of our proofs in this section and the correctness of known treewidth-based algorithms.

\iflong
Before we proceed to the three algorithms, we introduce some common notation that will be useful to describe the operation of dynamic programming algorithms using rank-width. For a node $t$ of a rooted rank decomposition $(T,\mu)$ of a graph $G$, let $S_t\subseteq V(G)$ be the set of vertices mapped by $\mu$ to leaves that $t$ separates from the root (i.e., $S_t$ contains the ``vertices below t'').
For disjoint vertex sets $S$ and $Q$ of $G$, two vertices $v$ and $w$ in $S$ are \emph{twins} with respect to $Q$ if $N(v)\cap Q=N(w)\cap Q$. A \emph{twin class} of $S$ with respect to $Q$ is a maximal subset of $S$ that consists of pairwise twins w.r.t.\ $Q$. Twin classes are a useful concept for designing dynamic programming algorithms that run on rank-decompositions, since there are at most \(2^{\rw(G)}\) twin classes of the set $S_t$ of vertices assigned in a subtree rooted at an arbitrary node $t\in V(T)$ with respect to the remaining vertices in the graph.
For each node \(t\) of \(T\) we enumerate these twin classes as \(R^t_1, R^t_2, \dotsc, R^t_z\), where \(z = 2^{\rw(G)}\) is a bound on the number of twin classes of \(S_t\) w.r.t.\ \(V(G) \setminus (X \cup S_t)\); in the case where there are only \(i<z\) twin classes, we let \(R^t_j= \emptyset\) for all \(i<j\leq z\).

For two children $t_1$ and $t_2$ of some node, we define $L_{t_1, t_2}$ as a matrix whose rows are indexed by the indices of twin classes of $S_{t_1}$ and columns are indexed by the indices of twin classes of $S_{t_2}$ such that for $L_{t_1, t_2}[j_1, j_2]=1$ if $R^{t_1}_{j_1}$ is complete to $R^{t_2}_{j_2}$, and $L_{t_1, t_2}[j_1, j_2]=0$ otherwise. For a child $t'$ of a node $t$, we define a function $U_{t', t}$ from the indices of twin classes of $S_{t'}$ to the indices of twin classes of $S_{t}$ such that for $U_{t', t}(j_1)=j_2$ if $R^{t'}_{j_1}$ is fully contained in $R^t_{j_2}$. This function is well defined because every twin class at $S_{t'}$ is fully contained in one of twin classes at $S_t$. Informally, $L_{t_1, t_2}$ specifies which edges are created between twin classes at $t$, while $U_{t_1}$ and $U_{t_2}$ specifies how twin-classes are ``reshuffled'' at $t$; note that all of these matrices can be pre-computed in polynomial time.

\subsection{Chromatic Number}\fi
\ifshort \smallskip \noindent \textbf{Chromatic Number. }\fi
In \textsc{Chromatic Number}, we are given a graph \(G\) and asked for the smallest number \(\chi(G)\) such that the vertex set of \(G\) can be properly colored using \(\chi(G)\) colors,
i.e., the smallest number \(\chi(G)\) such that \(V(G)\) can be partitioned into \(\chi(G)\) independent sets.
Our aim in this section is to solve a variant of \textsc{Chromatic Number} on graphs with a $k$-vertex modulator $X$ to $\cRc{c}$ where $X$ is precolored:

\begin{center}
\begin{mathproblem}{\textsc{$\cRc{c}$-Precoloring Extension}}[k]
	Instance: & A graph \(G\), a $k$-vertex modulator $X\subseteq V(G)$ to $\cRc{c}$ and a coloring of $X$.\\
	Task: & Compute the smallest number of colors required to extend the coloring of $X$ to a proper coloring of $G$.
\end{mathproblem}
\end{center}

\iflong\begin{theorem}\fi
\ifshort\begin{theorem}[$\clubsuit$]\fi
\label{thm:pre_chrom_number}
	\textsc{$\cRc{c}$-Precoloring Extension} can be solved in time $2^{\bigoh(k)} n^{\bigoh(1)}$. 
\end{theorem}

\ifshort
\begin{proof}[Proof Sketch]
	Let \(G\) be a graph together with a $k$-vertex modulator $X$ to $\cRc{c}$ and a proper coloring of $X$ by colors $[k]$; let $\text{col}_X$ be the set of colors assigned to at least one vertex in $X$. 
	For disjoint vertex sets $S$ and $Q$ of $G$, two vertices $v$ and $w$ in $S$ are \emph{twins} with respect to $Q$ if $N(v)\cap Q=N(w)\cap Q$. A \emph{twin class} of $S$ with respect to $Q$ is a maximal subset of $S$ that consists of pairwise twins w.r.t.\ $Q$.	
	
	Our starting point is a rooted rank-decomposition $(T, \mu)$ of $G-X$ of width at most $c$, which may be computed in time $\bigoh(n^3)$~\cite{HlinenyOum08}.
	On a high level, our algorithm will apply dynamic programming along $(T,\mu)$ where it will group colors together based on which twin classes they occur in (analogously as in the \XP algorithm for \textsc{Chromatic Number} parameterized by clique-width, due to Kobler and Rotics~\cite{KoblerR03}), but keep different (more detailed) records about the at most $k$ colors used in $X$.
	
	For each \(t \in V(T)\), let $S_t$ be the set of all vertices that are assigned to the descendants of $t$, and let $G_t:=G[S_t\cup X]$.
	By our definition of $(T,\mu)$, recall that $\rho_{G-X}(S_t)\leq c$ and that there are at most $z=2^c$ twin classes of $S_t$ w.r.t.\ $V(G)\setminus (X\cup S_t)$.
	We will refer to these twin classes as $R^t_1, R^t_2, \ldots, R^t_{z}$.
	
We are now ready to formally define the dynamic programming table $M_t$ that stores the information we require at a node $t$ of $T$. For $b_1, b_2, \ldots, b_k\subseteq [z]$ and $\SB d_Z\in [n] \SM Z\subseteq [z]\SE$ \todo{Possibly fix records (Eduard).
	}, we let $M_t(b_1, b_2, \ldots, b_k, \SB d_Z \SM Z\subseteq [z] \SE )=1$ if 
	there is a proper coloring of $G_t$ such that (1) for every $i\in [k]$, the color $i$ appears in twin classes in $\{R^t_j:j\in b_i\}$ and does not appear in other twin classes, and (2) for every $Z\subseteq [z]$, $d_Z$ is the number of colors from $\{k+1, k+2, \ldots, n\}$ that appear in twin classes of $\{R^t_j:j\in Z\}$ and do not appear in other twin classes. On the other hand, if no such proper coloring exists then we let $M_t(b_1, b_2, \ldots, b_k, \SB d_Z \SM Z\subseteq [z] \SE )=0$.
	
	The table $M_t$ will be filled in a leaf-to-root fashion. 
	Observe that by definition of $d_Z$'s, for distinct subsets $Z_1, Z_2$ of $[z]$,
	$d_{Z_1}$ and $d_{Z_2}$ count disjoint sets of colors.
	This provides an easy way to count the total number of colors used.
	Since all vertices in $G-X$ appear below the root node $r$, 
	the minimum number of colors required for a proper coloring of $G$ will be 
	the minimum value of  
$ \big| \text{col}_X \cup \SB i\in [k] \SM b_i\neq \emptyset \SE \big| + \sum_{Z\subseteq [z] }d_Z, $
	over all tuples $(b_1, b_2, \ldots, b_k, \SB d_Z \SM Z\subseteq [z]\SE)$ whose $M_r$ value is $1$.
\end{proof}
\fi

\iflong
\begin{proof}
	Let \(G\) be a graph together with a $k$-vertex modulator $X$ to $\cRc{c}$ and a proper coloring of $X$ by colors $\{1, 2, \ldots, k\}=[k]$; let $\text{col}_X$ be the set of colors assigned to at least one vertex in $X$. 
	Our starting point is a rooted rank-decomposition $(T, \mu)$ of $G-X$ of width at most $c$, which may be computed in time $\bigoh(n^3)$, using Theorem~\ref{thm:rankdecomp}.
	On a high level, our algorithm will proceed by dynamic programming along $(T,\mu)$ and group colors together based on which twin classes they occur in (analogously as in the \XP algorithm for \textsc{Chromatic Number} parameterized by clique-width, due to Kobler and Rotics~\cite{KoblerR03}), but it keeps different (more detailed) records about the at most $k$ colors used in $X$.
	
	Recall that, for any \(t \in V(T)\), $\rho_{G-X}(S_t) \leq c$ and that there are at most $z=2^c$ twin classes of $S_t$ w.r.t.\ $V(G)\setminus (X\cup S_t)$.
	
	\paragraph*{The Table.} We are now ready to formally define the dynamic programming table $M_t$ that stores the information we require at a node $t$ of $T$. For $b_1, b_2, \ldots, b_k\subseteq [z]$ and $\SB d_Z\in [n] \SM Z\subseteq [z]\SE$ \todo{Eduard: I don't like this being a set of numbers if each has its specific interpretation depending on $Z$, I would prefer it to be function $d: \mathcal{P}([z])\rightarrow [n]$, respectively $d: 2^{[z]}\rightarrow [n]$, where $\mathcal{P}([z])$, respectively $2^{[z]}$ is the power set (set of all subsets) of $[z]$ \\
	R: I think I agree with Eduard. However, while the current formulation is technically wrong (or at least abuses notation in a strange way), it doesn't hinder understanding of the proof---so if nobody has time to change this carefully in the proof, we can keep it.
	}, we let 
	\[M_t(b_1, b_2, \ldots, b_k, \SB d_Z \SM Z\subseteq [z] \SE )=1\] if 
	there is a proper coloring of $G_t$ such that 
	\begin{itemize}
		\item for every $i\in [k]$, the color $i$ appears in twin classes in $\{R^t_j:j\in b_i\}$ and does not appear in other twin classes, and
		\item for every $Z\subseteq [z]$, $d_Z$ is the number of colors from $\{k+1, k+2, \ldots, n\}$ that appear in twin classes of $\{R^t_j:j\in Z\}$ and do not appear in other twin classes;
	\end{itemize}
	on the other hand, if no such proper coloring exists, we let $M_t(b_1, b_2, \ldots, b_k, \SB d_Z \SM Z\subseteq [z] \SE )=0$.
	
	The table $M_t$ will be filled in a leaf-to-root fashion. 
	Observe that by definition of $d_Z$'s, for distinct subsets $Z_1, Z_2$ of $[z]$,
	$d_{Z_1}$ and $d_{Z_2}$ count disjoint sets of colors.
	This provides an easy way to count the total number of colors used.
	Since all vertices in $G-X$ appear below the root node \(\textit{ro}\), 
	the minimum number of colors required for a proper coloring of $G$ will be 
	the minimum value of  
	\[ \big| \text{col}_X \cup \SB i\in [k] \SM b_i\neq \emptyset \SE \big| + \sum_{Z\subseteq [z] }d_Z, \] 
	over all tuples $(b_1, b_2, \ldots, b_k, \SB d_Z \SM Z\subseteq [z]\SE)$ whose $M_\textit{ro}$ value is $1$.
	Thus, in order to complete the proof, it suffices to show how to fill in the table $M_t$ for each node $t\in V(T)$.
	
	\paragraph*{Leaf Nodes.} 	
	Consider a leaf node $t$, and assume that a vertex $v$ is assigned to $t$.
	To be a proper coloring, $v$ has to have a color that differs from the colors in $N_G(v)\cap X$.
	So, either we choose a color $i\in [k]$ that does not appear in $N_G(v)\cap X$, in which case we let $M_t$ assign the tuple ($b_i=1$ and all other values are $0$) to $1$, or
	we choose a color outside of $[k]$, in which case we let $M_t$ assign the tuple (all of $b_i$'s and $d_Z$'s are 0, except $d_{\{j\}}=1$ where $j$ is the label of the twin class containing $v$) to $1$. All other tuples are assigned a value of $0$ by $M_t$.

	\paragraph*{Internal Nodes.}
	Let us now consider an internal node $t$ with children $t_1$, $t_2$.
	To check whether $M_t(b^t_1, b^t_2, \ldots, b^t_k, \SB d^t_Z \SM Z\subseteq [z]\SE)=1$ for some tuple $\mathcal{T}=(b^t_1, b^t_2, \ldots, b^t_k, $ $\SB d^t_Z \SM Z\subseteq [z] \SE)$ at $t$, 
	we proceed as below, for all pairs of tuples $\mathcal{T}_i=(b^{t_i}_1, \ldots, b^{t_i}_k, $ $\SB d^{t_i}_Z \SM Z\subseteq [z] \SE)$ stored at $t_i$, $i\in [2]$.
	
	First, for the colors in $[k]$,  we have to check whether they are used in the twin classes in $R^{t_1}_{j_1}$ and $R^{t_2}_{j_2}$
	that are complete to each other (meaning that an edge will be created connecting vertices of the same color), and if yes then we discard this tuple and proceed to the next. More precisely, we check the following:
	\begin{itemize}
		\item (Compatibility 1) For every $i\in [k]$ with $b^{t_1}_i\neq \emptyset$ and $b^{t_2}_i\neq \emptyset$,  
		there are no $j_1\in b^{t_1}_i$ and $j_2\in b^{t_2}_i$ such that $L_{t_1, t_2}[j_1, j_2]=1$.
	\end{itemize}
	
	Second, we need to deal with the $d^t_Z$ components in our tables. Here, we branch on (``guess'')
	how many colors that appear in twin classes with indices in $d^{t_1}_{Z_1}$ are the same as the colors that appear in in twin classes with indices in $d^{t_2}_{Z_2}$ for all pairs $Z_1$ and $Z_2$.
	In particular, for all $Z_1, Z_2\subseteq [z]$, let $\tau(Z_1, Z_2)$ be a non-negative integer; this will correspond to the number of common colors in $d^{t_1}_{Z_1}$ and $d^{t_2}_{Z_2}$.
	Similar to (Compatibility 1) we have to check that for two complete twin classes, the same color does not appear at the same time.
	So, for each such mapping $\tau$, we check the following:
	\begin{itemize}
		\item (Compatibility 2) For every $Z_1, Z_2\subseteq [z]$ with $\tau(Z_1, Z_2)>0$,
		there are no $j_1\in Z_1$ and $j_2\in Z_2$ such that $L_{t_1, t_2}[j_1, j_2]=1$.
	\end{itemize}
	Observe that if Compatibility 2 would be violated, then $R^{t_1}_{j_1}$ and $R^{t_2}_{j_2}$ contain the same color while being complete, meaning that the result would not be a proper coloring. Hence if Compatibility 2 is violated, we discard and proceed to the next available branch (i.e., choice of $\tau$).
	
	Third, we check whether $\mathcal{T}_1$ and $\mathcal{T}_2$ will be combined into $\mathcal{T}$ with the function $\tau$.
	Recall that the function $U_{t_i, t}$ provides us with information about which set $R^{t}_{j'}$ will contain $R^{t_i}_j$. 
	We will use this function to rename our twin classes accordingly.
	When we perform this renaming, we also update $\tau$, $d^{t_1}_Z$, $d^{t_2}_Z$ 
	and obtain new functions $\tau'$, $\mathit{dm}^{t_1}_Z$, $\mathit{dm}^{t_2}_Z$, respectively.
	For example, if index $2$ were to be renamed to index $1$ at node $t_1$ and $\tau(\{1,2\}, \{3, 4\})=3$, 
	then $\tau'(\{1,2\}, \{3, 4\})$ becomes $0$ (since twin class $2$ is empty after renaming) and we apply $\tau'(\{1\}, \{3,4\})\leftarrow \tau(\{1\}, \{3,4\})+\tau(\{1,2\}, \{3,4\})$.
	\begin{itemize}
		\item (Renaming) For each $i\in \{1, 2\}$ and $j\in [z]$, we rename twin class $R^{t_i}_j$ to $\mathit{Rm}^{t_i}_{U_{t_i, t}(j)}$, and update 
		$\tau$, $d^{t_1}_Z$, $d^{t_2}_Z$ 
		to $\tau'$, $\mathit{dm}^{t_1}_Z$, $\mathit{dm}^{t_2}_Z$ accordingly.
	\end{itemize}
	Finally, for each $Z\subseteq [z]$ we check whether the number of colors in $\mathit{dm}^{t_1}_Z$ and $\mathit{dm}^{t_2}_Z$ correctly add up to $d^t_Z$ when taking $\tau$ into account. This is carried out by checking the following equality holds.
	\begin{align*}
	d^t_Z &= 
	\begin{aligned}[t]
	& \mathit{dm}^{t_1}_Z - \sum_{Z'\subseteq [z]} \tau'(Z, Z')  \\   
	& + \mathit{dm}^{t_2}_Z -   \sum_{Z'\subseteq [z]} \tau'(Z', Z) \\
	& + \sum_{Z_1\subseteq [z]} \big(\sum_{Z_2\subseteq [z], Z_1\cup Z_2=Z} \tau'(Z_1, Z_2)\big).
	\end{aligned}
	\end{align*} 
	Informally, the first row counts the number of colors that are unique to $\mathit{dm}^{t_1}_Z$, i.e., do not appear on any twin classes in $t_2$. The second row then does the same for $\mathit{dm}^{t_2}_Z$. To correctly determine the actual number of colors in $d^t_Z$ that we obtain by following $\tau$, we now need to add colors which occur in both $t_1$ and $t_2$ and which end up precisely in $Z$ -- that is what the third row counts.
	Summarizing, we assign $M_t(b^t_1, b^t_2, \ldots, b^t_k, \SB d^t_Z \SM Z\subseteq [z] \SE)=1$ if all the above conditions are satisfied in at least one branch, and use $0$ otherwise.
	
	We conclude the proof by analyzing the running time of the described algorithm. For a leaf node $t$, we can fill in the table in time $\mathcal{O}(k)$. 
	For an internal node $t$, there are at most $2^{zk}\cdot n^{2^z}$ different tuples $(b^t_1, b^t_2, \ldots, b^t_k, \SB d^t_Z \SM Z\subseteq [z]\SE)$. For each,
	(Compatibility 1) can be checked in time $\mathcal{O}(k)$.
	The number of possible functions $\tau$ is the same as the number of possible assignments of a number from $[n]$ to each pair $Z_1, Z_2\subseteq [z]$.
	Thus, it is bounded by $n^{2^{2z}}$. 
	Finally, updating $\tau$, $d^{t_1}_Z$, $d^{t_2}_Z$ 
	to $\tau'$, $\mathit{dm}^{t_1}_Z$, $\mathit{dm}^{t_2}_Z$ can be done in time $\bigoh(k)$.
	So, for a given coloring on $X$, we can find a minimum number of necessary colors in time $2^{zk}n^{\bigoh(1)}$.
\end{proof}		

\subsection{Hamiltonian Cycle}\label{subsec:ham_cycle}
\fi
\ifshort \smallskip \noindent \textbf{Hamiltonian Cycle. }\fi
In \textsc{Hamiltonian Cycle}, we are given an $n$-vertex graph $G$ and asked whether $G$ contains a cycle of length $n$ as a subgraph. Note that if we restrict $G$ to some subset of vertices $Y\subseteq V(G)$, then what remains from a Hamiltonian Cycle in $G$ is a set of paths that start and end in the neighborhood of $V(G)\setminus Y$. Hence, the aim of this section is to solve the following generalization of  \textsc{Hamiltonian Cycle}:

\begin{center}
	\begin{mathproblem}{\textsc{$\cRc{c}$-Disjoint Paths Cover}}[\(k\)]
		Instance: & A graph \(G\),  
		a $k$-vertex modulator $X\subseteq V(G)$ to $\cRc{c}$, and $m\le k$ pairs $(s_1,t_1),\ldots, (s_m,t_m)$ of vertices from $X$ with $s_i\neq t_i$ for all $i\in [m]$.\\
		Task: &  Decide whether there are internally vertex-disjoint paths $P_1, P_2,\ldots, P_m$ in $G$ such that $P_i$ is a path from $s_i$ to $t_i$ and every vertex in $G-X$ belongs to precisely one path in $P_1, P_2,\ldots, P_m$.
	\end{mathproblem}
\end{center}

\iflong\begin{theorem}\fi
\ifshort\begin{theorem}[$\clubsuit$]\fi
\label{thm:disjoint_paths_cover}
	\textsc{$\cRc{c}$-Disjoint Paths Cover} can be solved in time $2^{\bigoh(k)}n^{\bigoh(1)}$. 
\end{theorem}

\ifshort
\begin{proof}[Proof Sketch]
	Let \(G\) be a graph together with a $k$-vertex modulator $X$ to $\cRc{c}$ and $m$ pairs $(s_1,t_1),\ldots, $ $(s_m,t_m)$ of vertices from $X$. 
	Our starting point is once again a rooted rank-decomposition $(T, \mu)$ of $G-X$ of width at most $c$, which may be computed in time $\bigoh(n^3)$\ifshort~\cite{HlinenyOum08}\fi\iflong, using Theorem~\ref{thm:rankdecomp}\fi.
	We will obtain a fixed-parameter algorithm for checking the existence of such paths $P_1,\dots,P_m$ in $G$ by expanding the records used in Espelage, Gurski and Wanke's algorithm~\cite{EspelageGW03} for computing \textsc{Hamiltonian Cycle} parameterized by clique-width.	
	
	To follow partial solutions on each subgraph $G_t$, we consider certain generalizations of path-partitions of subgraphs of $G$. 
	For a subgraph $H$ of $G$, 
	an \emph{\(X\)-lenient path-partition} $\mathcal{P}$ of $H$
	is a collection of paths in \(H\) that are internally vertex-disjoint and share only endpoints in \(X\) such that $\bigcup_{P\in \mathcal{P}}V(P)=V(H)$.
	For convenience, we consider a path as an ordered sequence of vertices, 
	and for a path $P=v_1v_2 \cdots v_x$, we define $\ell(P)=v_1$ and $r(P)=v_x$.
	
	We proceed by introducing our dynamic programming table. 
	For each node $t$ of $T$, we use the following tuples $(D, \mathit{SP})$ as indices of the table.
	Let $D=\{d_{b_1, b_2}\in \{0, 1, \ldots, n\} \mid (b_1, b_2)\in [z]\times [z]\}$. 
	The integer $d_{b_1, b_2}$ will represent the number of paths in an \(X\)-lenient path-partition of $G_t$ that are fully contained in $G_t-X$ and whose endpoints are contained in $R^t_{b_1}$ and $R^t_{b_2}$.
	Let $\mathit{SP}$ be a set such that 
	\begin{itemize}
		\item for each $i\in \{1, \ldots, m\}$, $(i,0,x)$, $(0,i, x)$, $(i,i)$ with some $x\in [z]$ are the only possible tuples in $\mathit{SP}$, 
		\item each integer in $\{1, \ldots, m\}$ appears at most once as an $\ell$ among all tuples $(\ell, 0, p)$ or $(\ell, \ell)$ in $\mathit{SP}$, and similarly, 
		each integer in $\{1, \ldots, m\}$ appears at most once as an $r$ among all tuples $(0, r, p)$ or $(r, r)$ in $\mathit{SP}$.
	\end{itemize}
	In short, the tuple $(i,0,t)$ indicates the existence of a path starting in $s_i$ and ending at a vertex in $R^t_x$.
	Similarly, $(0,i,t)$ indicates the existence of a path starting in \(t_i\) and ending in \(R^t_x\).
	The tuple $(i,i)$ indicates the existence of a path starting in $s_i$ and ending in $t_i$.
	Note that there are at most $(n+1)^{z^2}$ possibilities for $D$. 
	For an element of $\mathit{SP}$, 
	there are $2mz+1$ possible elements in $\mathit{SP}$, and thus there are at most $2^{2kz+1}$ possibilities for $\mathit{SP}$.
	This implies that the number of possible tuples $(D, \mathit{SP})$ is bounded by $(n+1)^{z^2} 2^{2kz+1}$.
	
	We define a DP table $M_t$ such that $M_t(D, \mathit{SP})=1$ if 
	there is an \(X\)-lenient path-partition $\mathcal{P}=\mathcal{P}_1\uplus \mathcal{P}_2$ of $G_t$ such that 
	\begin{itemize}
		\item $\mathcal{P}_1$ is the subset of $\mathcal{P}$ that consists of all paths fully contained in $G_t-X$,
		\item for every $d_{b_1, b_2}\in D$, there are exactly $d_{b_1, b_2}$ distinct paths in $\mathcal{P}_1$ with endpoints in $R^t_{b_1}$ and $R^t_{b_2}$,
		\item for every \((\ell, r, p)\) or \((\ell, r) \in \mathit{SP}\),
		there is a unique path $P\in \mathcal{P}_2$ such that 
		\begin{itemize}
			\item if $\ell=i>0$, then $\ell(P)=s_i$, and 
			if $r=i>0$, then $r(P)=t_i$,
			\item if $\ell=0$, then $\ell(P)\in R^t_p$, and 
			if $r=0$, then $r(P)\in R^t_p$,
		\end{itemize}
	\end{itemize}
	In this case, we say that the \(X\)-lenient path-partition $\mathcal{P}$ is a partial solution with respect to $(D,\mathit{SP})$, and also 
	$(D,\mathit{SP})$ is a characteristic of $\mathcal{P}$.
	We define $\mathcal{Q}_t$ as the set of all tuples $(D, \mathit{SP})$ where $M_t(D, \mathit{SP})=1$.
	
	The table $M_t$ is filled in a leaf-to-root fashion.
	Since all vertices in $G-X$ appear below the root node \(\textit{ro}\), to decide whether there is a desired \(X\)-lenient path-partition,
	it suffices to confirm that there are $D$ and $\mathit{SP}$ such that $M_{\textit{ro}}(D, \mathit{SP})=1$, for every $d_{b_1, b_2}\in D$, $d_{b_1, b_2}=0$, and for every $i\in \{1, 2, \ldots, m\}$, $(i,i)\in \mathit{SP}$.

	The proof can be completed by describing a dynamic program to fill in the table $M_t$ for each node $t\in V(T)$ in a leaf-to-root fashion.
\end{proof}
\fi
\iflong
\begin{proof}
	Let \(G\) be a graph together with a $k$-vertex modulator $X$ to $\cRc{c}$ and $m$ pairs $(s_1,t_1),\ldots, (s_m,t_m)$ of vertices from $X$. 
	Our starting point is once again a rooted rank-decomposition $(T, \mu)$ of $G-X$ of width at most $c$, which may be computed in time $\bigoh(n^3)$, using Theorem~\ref{thm:rankdecomp}.
	We will obtain a fixed-parameter algorithm for checking the existence of such paths $P_1,\dots,P_m$ in $G$ by expanding the records used in Espelage, Gurski and Wanke's algorithm~\cite{EspelageGW03} for computing \textsc{Hamiltonian Cycle} parameterized by clique-width.

	
	To follow partial solutions on each subgraph $G_t$, we consider certain generalizations of path-partitions of subgraphs of $G$. 
	For a subgraph $H$ of $G$, 
	an \emph{\(X\)-lenient path-partition} $\mathcal{P}$ of $H$
	is a collection of paths in \(H\) that are internally vertex-disjoint and share only endpoints in \(X\) such that $\bigcup_{P\in \mathcal{P}}V(P)=V(H)$.
	For convenience, we consider a path as an ordered sequence of vertices, 
	and for a path $P=v_1v_2 \cdots v_x$, we define $\ell(P)=v_1$ and $r(P)=v_x$.
	
	\paragraph*{The Table.}
	For each node $t$ of $T$, we use the following tuples $(D, \mathit{SP})$ as indices of the table.
	Let $D=\{d_{b_1, b_2}\in \{0, 1, \ldots, n\} \mid (b_1, b_2)\in [z]\times [z]\}$. 
	The integer $d_{b_1, b_2}$ will represent the number of paths in an \(X\)-lenient path-partition of $G_t$ that are fully contained in $G_t-X$ and whose endpoints are contained in $R^t_{b_1}$ and $R^t_{b_2}$.
	Let $\mathit{SP}$ be a set such that 
	\begin{itemize}
		\item for each $i\in \{1, \ldots, m\}$, $(i,0,x)$, $(0,i, x)$, $(i,i)$ with some $x\in [z]$ are the only possible tuples in $\mathit{SP}$, 
		\item each integer in $\{1, \ldots, m\}$ appears at most once as an $\ell$ among all tuples $(\ell, 0, p)$ or $(\ell, \ell)$ in $\mathit{SP}$, and similarly, 
		each integer in $\{1, \ldots, m\}$ appears at most once as an $r$ among all tuples $(0, r, p)$ or $(r, r)$ in $\mathit{SP}$.
	\end{itemize}
	In short, the tuple $(i,0,t)$ will indicate the existence of a path starting in $s_i$ and ending in a vertex in $R^t_x$.
	Similarly, $(0,i,t)$ indicates the existence of a path starting in \(t_i\) and ending in \(R^t_x\).
	The tuple $(i,i)$ will indicate the existence of a path starting in $s_i$ and ending in $t_i$.
	%
	Note that there are at most $(n+1)^{z^2}$ possibilities for $D$. 
	For an element of $\mathit{SP}$, 
	there are $2mz+1$ possible elements in $\mathit{SP}$, and thus there are at most $2^{2kz+1}$ possibilities for $\mathit{SP}$.
	It implies that the number of possible tuples $(D, \mathit{SP})$ is bounded by $(n+1)^{z^2} 2^{2kz+1}$.
	
	We define a DP table $M_t$ such that $M_t(D, \mathit{SP})=1$ if 
	there is an \(X\)-lenient path-partition $\mathcal{P}=\mathcal{P}_1\uplus \mathcal{P}_2$ of $G_t$ such that 
	\begin{itemize}
		\item $\mathcal{P}_1$ is the subset of $\mathcal{P}$ that consists of all paths fully contained in $G_t-X$,
		\item for every $d_{b_1, b_2}\in D$, there are exactly $d_{b_1, b_2}$ distinct paths in $\mathcal{P}_1$ whose endpoints are contained in $R^t_{b_1}$ and $R^t_{b_2}$,
		\item for every \((\ell, r, p)\) or \((\ell, r) \in \mathit{SP}\),
		there is a unique path $P\in \mathcal{P}_2$ such that 
		\begin{itemize}
			\item if $\ell=i>0$, then $\ell(P)=s_i$, and 
			if $r=i>0$, then $r(P)=t_i$,
			\item if $\ell=0$, then $\ell(P)\in R^t_p$, and 
			if $r=0$, then $r(P)\in R^t_p$,
		\end{itemize}
	\end{itemize}
	In this case, we say that the \(X\)-lenient path-partition $\mathcal{P}$ is a partial solution with respect to $(D,\mathit{SP})$, and also 
	$(D,\mathit{SP})$ is a characteristic of $\mathcal{P}$.
	We say that paths in $\mathcal{P}_1$ are normal paths, and paths in $\mathcal{P}_2$ are special paths.
	The value $M_t(D, \mathit{SP})$ is $0$ if there is no such \(X\)-lenient path-partition of $G_t$.
	We define $\mathcal{Q}_t$ as the set of all tuples $(D, \mathit{SP})$ where $M_t(D, \mathit{SP})=1$.
	
	The table $M_t$ will be filled in a leaf-to-root fashion.
	Since all vertices in $G-X$ appear below the root node \(\textit{ro}\), to decide whether there is a desired \(X\)-lenient path-partition,
	it suffices to confirm that there are $D$ and $\mathit{SP}$ such that 
	\begin{itemize}
		\item $M_{\textit{ro}}(D, \mathit{SP})=1$,
		\item for every $d_{b_1, b_2}\in D$, $d_{b_1, b_2}=0$,
		\item for every $i\in \{1, 2, \ldots, m\}$, $(i,i)\in \mathit{SP}$.
	\end{itemize}

	Thus, in order to complete the proof, it suffices to show how to fill in the table $M_t$ for each node $t\in V(T)$.
	
	\paragraph*{Leaf Nodes.}
	First consider a leaf node $t$. Let $J =  \SB i\in [m] \SM \{s_i,t_i\}\in E(G) \SE$. For every pair $i\in J$, we can branch on whether a $s_i$-$t_i$ path will be only the edge $\{s_i,t_i\}$, or it will contain also some vertices from $G-X$. For $I\subseteq J$, let us denote by $\mathit{SP}_{I}$ the set $\SB (i,i) \SM i\in I \SE$. Moreover, let a vertex $v$ be assigned to $t$, and $x$ be the index of the twin class $\{v\}$. 
	If it is adjacent to $s_i$ or $t_i$ for some $i$, 
	then we may select it as a starting point or an ending point of a desired path or both. 
	So, for every tuple $(D,\mathit{SP})$ such that $d_{b_i, b_j}=0$ for all $b_i, b_j\in [z]$, and $\mathit{SP}= \mathit{SP}_I\cup \mathit{SP}_i$, where $I\subseteq J\setminus \{i\}$ and 
	\[\mathit{SP}_i = \left\{ \begin{array}{ll}
	\{(i, 0, x)\} & \qquad \textrm{if $v\in N_G(s_i)$}\\
	\{(0,i,x)\} & \qquad  \textrm{if $v\in N_G(t_i)$}\\
	\{(i,i)\}& \qquad  \textrm{if both holds,}
	\end{array} \right.
	\]
	we give $M_t(D, \mathit{SP})=1$. Note, that if $v\in N_G(s_i)\cap N_G(t_i)$, then there is a tuple for each of the three possibilities for $\mathit{SP}_i$.
	Also, we could just keep it as a path whose endpoints are not linked to $X$ in the final solution.
	In this case, for a tuple $(D,\mathit{SP})$ where 
	\begin{itemize}
		\item $d_{x,x}=1$ and all other $d_{y,y'}$'s have values $0$,
		\item $\mathit{SP}=\mathit{SP}^I$ for some $I\subseteq J$,
	\end{itemize} 
	we give $M_t(D, \mathit{SP})=1$.
	Other tuples get the value $0$.

	\paragraph*{Internal Nodes.}
	Let $t$ be an internal node, and let $t_1, t_2$ be the children of $t$.
	We assume that $\mathcal{Q}_{t_1}$ and $\mathcal{Q}_{t_2}$ were already computed, and 
	we want to generate $\mathcal{Q}_t$ from these sets.
	
	For all pairs of tuples $(D^1, \mathit{SP}^1)\in \mathcal{Q}_{t_1}$ and $(D^2, \mathit{SP}^2)\in \mathcal{Q}_{t_2}$, 
	we will generate all possible tuples $(D, \mathit{SP})$ such that for an \(X\)-lenient path-partition $\mathcal{P}$ with respect to $(D, S)$, 
	its restriction to $G_{t_i}$ has $(D^i, \mathit{SP}^i)$ as a characteristic.
	
	We first test basic conditions that the two partial solutions can be a partial solution in $G_t$.
	The starting point or an endpoint of a special path in the final solution should start with one vertex in $G-X$.
	So, the following condition has to be satsified.
	\begin{itemize}
		\item (Special terminal) Each integer in $\{1, \ldots, m\}$ appears at most once as an $\ell$ among all tuples $(\ell, 0, p)$ or $(\ell, \ell)$ in $\mathit{SP}^1\cup \mathit{SP}^2$, and similarly, 
		each integer in $\{1, \ldots, m\}$ appears at most once as an $r$ among all tuples $(0, r, p)$ or $(r, r)$ in $\mathit{SP}^1\cup \mathit{SP}^2$.
	\end{itemize}
	If (Special terminal) condition is not satisfied, then we skip this pair of tuples.
	In the rest, we assume that (Special terminal) condition is satisfied.
	
	Secondly, we rename twin classes to consider both partial solutions together.
	\begin{itemize}
		\item (Renaming) For each $i\in \{1, 2\}$ and $j\in [z]$, we rename twin class $R^{t_i}_j$ to $\mathit{Rm}^{t_i}_{U_{t_i, t}(j)}$, and update 
		$D^1, D^2, \mathit{SP}^1, \mathit{SP}^2$ to $\mathit{Dm}^1, \mathit{Dm}^2, \mathit{SPm}^1, \mathit{SPm}^2$ accordingly.
	\end{itemize}
	For example, if index $2$ were to be renamed to index $1$ at node $t_1$ and $d^1_{2,3}=3$, 
	then $\mathit{dm}^1_{2,3}$ in $\mathit{Dm}^1$ becomes $0$ and apply $\mathit{dm}^1_{1,3}$ in $\mathit{Dm}^1$ to $d^1_{1,3}+d^1_{2,3}$.
	Similarly, if there was a tuple $(i,0,2)$ in $\mathit{SP}^1$, then we add a tuple $(i,0, 1)$ to $\mathit{SPm}^1$.
	Because this simply change the indices of twin classes, (Special terminal) condition is still valid.
	After then we take the sum of $(\mathit{Dm}^1, \mathit{SPm}^1)$ and $(\mathit{Dm}^2, \mathit{SPm}^2)$ which correspond to the disjoint union of partial solutions with respect to those tuples.
	More precisely, we take that 
	\begin{itemize}
		\item $D^{\mathit{begin}}=\{d^{\mathit{begin}}_{x,y} | x,y\in [z]\}$ where $d^{\mathit{begin}}_{x,y}=\mathit{dm}^1_{x,y}+\mathit{dm}^2_{x,y}$ for all $x,y\in [z]$ and $\mathit{dm}^1_{x,y}\in \mathit{Dm}^1$ and $\mathit{dm}^2_{x,y}\in \mathit{Dm}^2$, and 
		\item $\mathit{SP}^{\mathit{begin}}=\mathit{SPm}^1\cup \mathit{SPm}^2$.
	\end{itemize}
	Because of the condition (Special terminal), 
	no integer in $\{1, \ldots, m\}$ appears twice in $\mathit{SP}^{\mathit{begin}}$. Therefore, $(D^{\mathit{begin}}, \mathit{SP}^{\mathit{begin}})$ is a possible tuple for $G_t$, which corresponds to 
	exactly the disjoint union of partial solutions with respect to $(D^1, \mathit{SP}^1)$ and $(D^2, \mathit{SP}^2)$. 
	We put it to $\mathcal{Q}_t$.
	
	Next, we consider to merge two paths in an \(X\)-lenient path-partition, using edges between $V(G_{t_1})\setminus X$ and $V(G_{t_2})\setminus X$.
	There are three types of merging operations: merging two normal paths, merging a normal path and a special path, and merging two special paths.
	\begin{itemize}
		\item (Merging two normal paths) Suppose there are $x_1, y_1, x_2, y_2\in [z]$ with $x_1'=U_{t_1,t}(x_1)$, $y_1'=U_{t_1,t}(y_1)$, $x_2'=U_{t_2,t}(x_2)$, $y_2'=U_{t_2,t}(y_2)$ such that $R^{t_1}_{x_1}$ is complete to $R^{t_2}_{x_2}$ and $d^1_{x_1, y_1}>0$ and $d^2_{x_2, y_2}>0$. 
		Then we subtract 1 from each of $d^{\mathit{begin}}_{x_1', y_1'}$ and $d^{\mathit{begin}}_{x_2', y_2'}$ and add 1 to $d^{\mathit{begin}}_{y_1', y_2'}$.
		\item (Merging a normal path and a special path) 	Suppose there are $x, x_2, y_2\in [z]$ with $x'=U_{t_1,t}(x)$, $x_2'=U_{t_2,t}(x_2)$, $y_2'=U_{t_2,t}(y_2)$ such that $R^{t_1}_{x}$ is complete to $R^{t_2}_{x_2}$ and $(i,0,x)\in \mathit{SP}^1$ and $d^2_{x_2, y_2}>0$. 
		Then we subtract 1 from $d^{\mathit{begin}}_{x_2', y_2'}$ and replace $(i,0, x)$ with $(i,0, y_2')$ in $\mathit{SP}^{\mathit{begin}}$. 
		The symmetric operation is also applied.
		\item (Merging two special paths) 	Suppose there are $x, y\in [z]$ with $x'=U_{t_1,t}(x)$, $y'=U_{t_2,t}(y)$ such that $R^{t_1}_{x}$ is complete to $R^{t_2}_{y}$ and $(i,0,x)\in \mathit{SP}^1$ and  $(0,i,y)\in \mathit{SP}^2$.
		Then we remove $(i,0, x')$ and $(0,i, y')$ from $\mathit{SP}^{\mathit{begin}}$ and add $(i,i)$ to $\mathit{SP}^{\mathit{begin}}$.
	\end{itemize}
	To keep track of the information on how many edges are used, we also have to change the information from $(D^1, \mathit{SP}^1)$ and $(D^2, \mathit{SP}^2)$.
	We recursively apply this process of merging two paths until we can no longer get a tuple $(D', \mathit{SP}')$ that is not yet in $\mathcal{Q}_t$.
	Because the number of possible tuples $(D, \mathit{SP})$ is bounded by $(n+1)^{z^2} 2^{2kz+1}$, 
	this process can be done in time $\mathcal{O}((n+1)^{z^2} 2^{2kz+1})$ for a fixed pair of tuples $(D^1, \mathit{SP}^1)$ and $(D^2, \mathit{SP}^2)$.
	Considering all pairs of tuples in $\mathcal{Q}_{t_1}$ and $\mathcal{Q}_{t_2}$, 
	we can update $\mathcal{Q}_t$ in time $\mathcal{O}((n+1)^{3z^2} 2^{6kz+3})$.
	Over all, we can update $\mathcal{Q}_t$ for all $t$ in time $\mathcal{O}((n+1)^{3z^2+1} 2^{6kz+3})$.
	As $z$ is constant, this problem can be solved in time $2^{\bigoh(k)}n^{\bigoh(1)}$.
\end{proof}

\subsection{Max-Cut}
\fi
\ifshort \smallskip \textbf{Max-Cut. }\fi
The third problem we consider is \textsc{Max-Cut}, where we are given an integer $\ell$ together with an $n$-vertex graph $G$ and asked whether $V(G)$ can be partitioned into sets $V_1$ and $V_2$ such that the number of edges with precisely one endpoint in $V_1$ (called the \emph{cut size}) is at least $\ell$.

\begin{center}
	\begin{mathproblem}{\textsc{$\cRc{c}$-Max-Cut Extension}}[k]
		Instance: & A graph \(G\), a $k$-vertex modulator $X\subseteq V(G)$ to $\cRc{c}$, $s\subseteq X$, and $\ell \in \Nat$.\\
		Task: & Is there a partition of $V(G)$ into sets $V_1$ and $V_2$ such that $X\cap V_1= s$ and the number of edges between $V_1$ and $V_2$ is at least $\ell$.
	\end{mathproblem}
\end{center}

\iflong\begin{theorem}\fi
\ifshort\begin{theorem}[$\clubsuit$]\fi
\label{thm:modulator_max_cut}
	\textsc{$\cRc{c}$-Max-Cut Extension} can be solved in polynomial time. 
\end{theorem}
\iflong
\begin{proof}
	Let \(G\) be a graph together with a \(k\)-vertex modulator \(X\) to \(\cRc{c}\) and a set $s\subseteq X$. Since, we fix the partition of $X$ beforehand, we can assume that $X$ is an independent set and only add the number of edges between $s$ and $X\setminus s$ at the end. 
	We can use Theorem~\ref{thm:rankdecomp} to obtain a rank-decomposition \((T, \mu)\) of \(G - X\) of width at most \(c\) in time in \(\bigoh(n^3)\).
	
	The general idea is similar to the one for the \XP algorithm for \textsc{Max-Cut}~\cite{GanianHO13}:
	We determine, via a dynamic programming table, how large the largest cut in \(G_t\) can be if we assume a certain number of vertices of each twin class to be in the same side of the cut as \(s\).
	
	\paragraph*{The Table.} The dynamic programming table consists of entries for each node \(t\) of \(T\),
	each of which in turn describe for any tuple \((c_1, \dotsc, c_{z}) \in [\vert R^t_1\vert] \times \dotsc [\vert R^t_{z} \vert]\) what the largest possible cut in \(G_t\) is such that \(c_i\) vertices of \(R^t_i\) are in the same side of the cut as \(s\).
	Formally,
	\[
	M_t(c_1, \dotsc, c_{z}) = \max_{\substack{s \subseteq C \subseteq V(G)\\ \forall 1 \leq i \leq z \ \vert C \cap R^t_i \vert = c_i}} \vert \{e \in E(G) \mid e \cap C \neq \emptyset \land e \cap (V(G) \setminus C) \neq \emptyset\} \vert.
	\]
	
	\paragraph*{Leaf Nodes.}
	Actually, the only adaptation that has to be made for the consideration of \(s \subseteq X\) concerns the leaf nodes.
	Consider a leaf node \(t\) and let \(v = \mu^{-1}(t)\).
	There are two possibilities for a cut \(C \supseteq s\) of \(G_t\):
	Either we include \(v\) into the side of the cut containing \(s\),
	this corresponds to defining \(M_t(1) = \vert N_G(v) \cap (X \setminus s) \vert\),
	or we include \(v\) into the side of the cut not containing \(s\),
	this corresponds to defining \(M_t(0) = \vert N_G(v) \cap X \vert\).
	
	\paragraph*{Internal Nodes.}
	Let us now consider an internal node \(t\) with children \(t_1\) and \(t_2\).
	A cut of \(G_t\) then always is the union of a cut in \(G_{t_1}\) and a cut in \(G_{t_2}\).
	The edges that contribute to the value of these cuts within \(G_{t_1}\) and \(G_{t_2}\) respectively still contribute to the value in \(G_t\).
	However, edges with an endpoint in \(V(G_{t_1})\) and an endpoint in \(V(G_{t_2})\) are not taken into account in \(G_{t_1}\) and \(G_{t_2}\).
	The number of these edges are not dependant on the specific cutsets within \(G_{t_1}\) and \(G_{t_2}\), but the size of their intesection with each of the twin classes.
	Thus, each entry of the dynamic programming table at \(t\) is given rise to by a pair of two entries \(M_{t_1}(c_1, \dotsc, c_{z_{t_1}})\)
	and \(M_{t_2}(d_1, \dotsc, d_{z_{t_2}})\) of the dynamic progamming table at \(t_1\) and \(t_2\), in the following way:
	\begin{align*}
	& M_t(c_{U_{t_1, t}(1)} + d_{U_{t_2, t}(1)}, \dotsc, c_{U_{t_1, t}(z_{t_1})} + d_{U_{t_2, t}(z_{t_2})})\\
	& = \ \begin{aligned}[t]
	& M_{t_1}(c_1, \dotsc, c_{z_{t_1}}) + M_{t_2}(d_1, \dotsc, d_{z_{t_2}})\\
	& + \sum_{i = 1}^{z} \vert R^t_{U_{t_1, t}(i)} \vert \cdot \vert R^t_{U_{t_2, t}(i)} \vert - c_{U_{t_1, t}(i)} \cdot d_{U_{t_2, t}(i)},
	\end{aligned}
	\end{align*}
	for certain \(c_1, \dotsc, c_{z_{t_1}}, d_1, \dotsc, d_{z_{t_2}}\).
	To determine the correct \(c_1, \dotsc, c_{z_{t_1}}, d_1, \dotsc, d_{z_{t_2}}\) we can take the maximum over all possible choices.
	
	We conclude the proof by analizing the running time of the described algorithm.
	For a leaf node \(t\), we can fill the table in time in \(\bigoh(k)\).
	For an internal node \(t\), there are at most \(n^{z} \leq n^{2^c}\) different tuples \((c_1, \dotsc, c_{z})\) and at most \(n^{z} \leq n^{2^c}\) possibilities to write each of the \(c_i\) as a 2-sum.
\end{proof}
\fi

\section{Algorithmic Applications of $\cRc{c}$-Treewidth}
\newcommand{\rec}{\delta}

In this section, we show that \textsc{Chromatic Number}, \textsc{Hamiltonian Cycle} and \textsc{Max-Cut} are \FPT\ parameterized by $\cRc{c}$-treewidth.
As our starting point, recall that each of these problems admits a fixed-parameter algorithm when parameterized by treewidth which is based on leaf-to-root dynamic programming along the nodes of a nice tree decomposition. Notably, the algorithms are based on defining a certain \emph{record} $\rec(P,Q)$ (for vertex sets $P$, $Q$) such that $\rec(B_t,Y_t)$ captures all the relevant information required to solve the problem on $G[Y_t]$ and to propagate this information from a node $t$ to its parent. The algorithms compute these records on the leaves of the tree decomposition by brute force, and then dynamically update these records while traversing a nice tree decomposition towards the root; once the record $\rec(B_r,Y_r)$ is computed for the root $r$ of the decomposition, the algorithm outputs the correct answer. 
\iflong We refer to the standard textbooks for a detailed description of dynamic programming along nice tree decompositions~\cite{CyganFominKowalikLokshtanovMarxPilipczukPilipcukSaurabh13,DowneyFellows13} and to Subsections~\ref{sub:chrom}--\ref{sub:maxcut} for an overview of the definition of the records used for the target problems.
\fi

Our general strategy for solving these problems will be to replicate the records employed by the respective dynamic programming algorithm $\mathbb{A}$ used for treewidth, but \emph{only for the nice $\cRc{c}$-tree decomposition of the torso} of the input graph $G$. Recall that aside from the ``standard'' simple leaf nodes, nice $\cRc{c}$-tree decompositions also contain boundary leaf nodes, which serve as separators between the torso and a connected component $C$ with rank-width at most $c$. For $\mathbb{A}$ to work correctly with the desired runtime, we need to compute the record for each boundary leaf node using a subprocedure that exploits the bounded rank-width of $C$; in particular, we will see that this amounts to solving the problems defined in Section~\ref{sec:modulator_algorithms}.
Before proceeding to the individual problems, we provide a formalization and proof for the general ideas outlined above.

\ifshort
\begin{lemma}
\label{lem:dynprog}
Let $\pP$ be a graph problem which can be solved via a fixed-parameter algorithm $\mathbb{A}$ parameterized by treewidth, where $\mathbb{A}$ runs in time $f(k')\cdot n'^a$ and operates by computing a certain record $\rec$ in a leaves-to-root fashion along a provided nice width-$k'$ tree decomposition of the $n'$-vertex input graph. 

Let $\pQ$ be obtained from $\pP$ by receiving the following additional information in the input: (1) a nice \(\cRc{c}\)-tree decomposition \((X, T, \{B_t \mid t \in V(T)\})\) of width $k$ for the input $n$-vertex graph $G$, and (2) for each boundary leaf node \(t\) corresponding to the neighborhood of a connected component \(C\) of \(G[X]\), the record $\rec(B_t,B_t\cup C)$.

Then, $\pQ$ can be solved in time $f(k)\cdot n^a$.
\end{lemma}
\fi
\iflong
\begin{lemma}
\label{lem:dynprog}
Let $\pP$ be a graph problem which can be solved via a fixed-parameter algorithm $\mathbb{A}$ parameterized by treewidth, where $\mathbb{A}$ runs in time $f(k')\cdot n'^a$ and operates by computing a certain record $\rec$ in a leaves-to-root fashion along a provided nice width-$k'$ tree decomposition of the $n'$-vertex input graph. 

Let $\pQ$ be obtained from $\pP$ by receiving the following additional information in the input:
\begin{itemize}
\item a nice \(\cRc{c}\)-tree decomposition \((X, T, \{B_t \mid t \in V(T)\})\) of width $k$ for the input $n$-vertex graph $G$, and 
\item for each boundary leaf node \(t\) corresponding to the neighborhood of a connected component \(C\) of \(G[X]\), the record $\rec(B_t,B_t\cup C)$.
\end{itemize}
Then, $\pQ$ can be solved in time $f(k)\cdot n^a$.
\end{lemma}
\begin{proof}
Consider the algorithm $\mathbb{B}$ which computes all the records $\rec$ for simple leaf nodes by calling the respective subroutine used in $\mathbb{A}$, uses the records $\rec(B_t,B_t\cup C)$ for the boundary leaf nodes, and then proceeds to compute the records in a dynamic leaves-to-root fashion in exactly the same way as $\mathbb{A}$. It is clear that $\mathbb{B}$ terminates in the claimed runtime.

To argue correctness, assume for a contradiction that $\mathbb{B}$ outputs incorrectly. Now consider the tree decomposition $(T',\{B'_t \mid t \in V(T')\}))$ of $G$ obtained by attaching, to each boundary leaf $t_C$ in $T$ (where $B_{t_C}=N(C)$ for some connected component $C$ in $G[X]$), a new leaf whose bag contains \(N(C) \cup C\).
While the width of such a decomposition may be linear in the number of vertices, $\mathbb{A}$ must still output the correct solution, and observe that the records computed by $\mathbb{A}$ on $(T',\{B'_t \mid t \in V(T')\}))$ must precisely match those computed by $\mathbb{B}$ on \((X, T, \{B_t \mid t \in V(T)\})\). Hence, we would in this case conclude that $\mathbb{A}$ also outputs incorrectly, a contradiction.
\end{proof}
\fi

\iflong
\subsection{Chromatic Number}\label{sub:chrom}\fi
\ifshort \smallskip \noindent \textbf{Chromatic Number. }\fi
\textsc{Chromatic Number} is \W{1}-hard parameterized by rank-width~\cite{FominGLS09} but can be solved in time $2^{\bigoh(\tw(G)\cdot \log \tw(G))}\cdot n$ on $n$-vertex graphs when a minimum-width tree-decomposition is provided with the input~\cite{JansenS97};
\ifshort moreover, it is known that this runtime is essentially tight~\cite{LokshtanovMS18}.\fi
\iflong moreover, it is known that this runtime is essentially tight, since under ETH the problem cannot be solved in time $2^{o(\tw(G)\cdot \log \tw(G))}\cdot n^{\bigoh(1)}$~\cite{LokshtanovMS18}.
\fi

It is well known that the chromatic number is at most $\tw(G)+1$\iflong~(this can be observed, e.g., by greedily coloring a vertex when it is introduced in the nice tree decomposition)\fi. One possible way of defining records in order to achieve a runtime of $2^{\bigoh(\tw(G)\cdot \log \tw(G))}\cdot n$ is to track, for each proper coloring of vertices in a bag $B_t$, the minimum number of colors required to extend such a coloring to $Y_t$~\cite{JansenS97}.
Formally, let $S_t$ be the set of all colorings of $B_t$ with colors $[\tw(G)+1 ]$, and let $\reca(B_t,Y_t) : S_t \to \mathbb{Z}$ be defined as follows:
\begin{itemize}
\vspace{-0.2cm}
\item $\reca(B_t,Y_t)(s)=-1$ if $s$ is not a proper coloring of $G[B_t]$.
\item $\reca(B_t,Y_t)(s)=q$ if $q$ is the minimum number of colors used by any proper coloring of $G[Y_t]$ which extends $s$. 
\vspace{-0.2cm}
\end{itemize}


Using Theorem~\ref{thm:pre_chrom_number}, we can compute such $\reca(B_t,Y_t)(s)$ for every proper coloring $s$ of $B_t$. Hence, combining Lemma~\ref{lem:dynprog} and Theorem~\ref{thm:pre_chrom_number}, we obtain:

\begin{theorem}\label{thm:chrom_number}
	\textsc{Chromatic Number} can be solved in time $2^{\bigoh(k\log(k))}\cdot n^{\bigoh(1)}$ if a nice \(\cRc{c}\)-tree decomposition of width $k$ is provided on the input. 
\end{theorem}


\iflong
\subsection{Hamiltonian Cycle}\label{sub:ham}\fi
\ifshort \smallskip \noindent \textbf{Hamiltonian Cycle. }\fi
 \textsc{Hamiltonian Cycle} is \W{1}-hard parameterized by rank-width~\cite{FominGLS09} but can be solved in time $2^{\bigoh(\tw(G)\cdot \log \tw(G))}\cdot n$ on $n$-vertex graphs when a minimum-width tree-decomposition is provided with the input via standard dynamic programming\iflong~(see, e.g., an introduction to advances made for the problem by Ziobro and Pilipczuk~\cite{ZiobroP18})\fi. 
This algorithm can be improved to run in time $2^{\bigoh(\tw(G))}\cdot n)$ by applying the advanced \emph{rank-based approach} of Cygan, Kratsch and Nederlof~\cite{CyganKN18} to prune the number of records.
To simplify our exposition, here we focus on extending the standard dynamic programming algorithm which yields a slightly super-exponential runtime.

One possibility for defining the records for \textsc{Hamiltonian Cycle} is to track all possible ways one can cover $Y_t$ by paths that start and end in $B_t$ (intuitively, this corresponds to what remains of a hypothetical solution if we ``cut off'' everything above $Y_t$)~\cite{DowneyFellows13}.
Formally, let \({{B_t}^{\diamond}}\) be defined as follows:
\begin{itemize}
\vspace{-0.2cm}
\item if $|B_t|>2$, then ${B_t}^{\diamond}$ is the set of graphs with at most $|B_t|$ edges and degree at most $2$ over vertex set $B_t$;
\item if $|B_t|=2$, then ${B_t}^{\diamond}$ contains three (multi)graphs over vertex set $B_t$: the edgeless graph, the graph with one edge, and the multigraph with two edges and no loops;
\item if $|B_t|=1$, then ${B_t}^{\diamond}$ contains an edgeless graph and a graph with a single loop, both over the single vertex in $B_t$;
\item if $|B_t|=0$, then ${B_t}^{\diamond}=\{$YES, NO$\}$.
\vspace{-0.2cm}
\end{itemize}

We let $\recb(B_t,Y_t): {{B_t}^{\diamond}} \to \{0, 1\}$, where for \(Q \in {{B_t}^{\diamond}}\) we set $\recb(B_t,Y_t)(Q)=1$ if and only if there exists a set $P$ of paths in \(G[Y_t]\) and a bijection that maps each $(v_1,\dots,v_\ell)\in P$ to an edge $(v_1,v_\ell)\in E(Q)$ such that each vertex $v\in G[Y_t\setminus B_t]$ is contained in precisely one path in $P$. In the special case where $B_t=\emptyset$, our records explicitly state whether $G[Y_t]$ contains a Hamiltonian cycle or not.

\iflong
As before, we can now shift our attention to the problem of computing our records in boundary leaf nodes.
If $|B_t|\le 1$, then, because $B_t$ is a separator in $G$, either $Y_t=B_t$, in which case $\recb(B_t,Y_t)$ is trivial to compute, or $\recb(B_t,Y_t)$ can be filled by simply solving \textsc{Hamiltonian cycle} on $G[Y_t]$ in polynomial time using known algorithms~\cite{GanianHO13}. 
In all other cases, we loop over all of the at most $k^{2k}$-many graphs $Q\in {{B_t}^{\diamond}}$ and for each such $Q$ we need to check whether $G[Y_t]-B_t$ can be covered by internally vertex-disjoint paths connecting the pairs of vertices in $B_t$ that form the endpoints of the edges in $Q$. Moreover, note that in this case $Q$ does not contains any loops. 
Hence, this is precisely the \textsc{$\cRc{c}$-Disjoint Paths Cover} problem defined in Subsection~\ref{subsec:ham_cycle} and combining Theorem~\ref{thm:disjoint_paths_cover} and Lemma~\ref{lem:dynprog}, we obtain:
\fi
\ifshort
As before, we can now shift our attention to the problem of computing our records in boundary leaf nodes.
We do so by looping over all of the at most $k^{2k}$-many graphs $Q\in {{B_t}^{\diamond}}$; for each such $Q$ we check whether $G[Y_t]-B_t$ can be covered by internally vertex-disjoint paths connecting the pairs of vertices in $B_t$ that form the endpoints of the edges in $Q$. 
Hence, we are left with the \textsc{$\cRc{c}$-Disjoint Paths Cover} problem. From Theorem~\ref{thm:disjoint_paths_cover} and Lemma~\ref{lem:dynprog}, we obtain:
\fi

\begin{theorem}\label{thm:hamiltonian_cycle}
	\textsc{Hamiltonian Cycle} can be solved in time $2^{\bigoh(k\log(k))}\cdot n^{\bigoh(1)}$ if a nice \(\cRc{c}\)-tree decomposition of width $k$ is provided on the input. 
\end{theorem}

\iflong
\subsection{Max-Cut}	\label{sub:maxcut}\fi
\ifshort \noindent \smallskip \textbf{Max-Cut. }\fi
\textsc{Max-Cut} is another problem that is \W{1}-hard parameterized by rank-width~\cite{FominGLS14} but admits a simple fixed-parameter algorithm parameterized by treewidth -- notably, it can be solved in time $2^{\bigoh(\tw(G))}\cdot n$ on $n$-vertex graphs when a minimum-width tree-decomposition is provided with the input via standard dynamic programming~\cite{CyganFominKowalikLokshtanovMarxPilipczukPilipcukSaurabh13,DowneyFellows13}. 

The simplest way of defining the records for \textsc{Max-Cut} is to keep track of all possible ways the bag $B_t$ can be partitioned into $V_1$ and $V_2$, and for each entry in our table we keep track of the maximum number of crossing edges in $Y_t$ compatible with that entry. Formally, let $\recc(B_t,Y_t):2^{B_t}\rightarrow \Nat_0$, 
where for each $s\in 2^{B_t}$ it holds that $\recc(B_t,Y_t)(s)$ is the maximum cut size that can be achieved in $G[Y_t]$ by any partition $(V_1, V_2)$ satisfying $V_1\cap B_t=s$.
As before, from Theorem~\ref{thm:modulator_max_cut} and Lemma~\ref{lem:dynprog}, we obtain:

\begin{theorem}\label{thm:max_cut}
	\textsc{Max-Cut} can be solved in time $2^{k}\cdot n^{\bigoh(1)}$, if a nice \(\cRc{c}\)-tree decomposition of width $k$ is provided on the input. 
\end{theorem}

\section{Concluding Remarks}
While the technical contribution of this paper mainly focused on $\cRc{c}$-treewidth, a parameter that allows us to lift fixed-parameter algorithms parameterized by treewidth to well-structured dense graph classes, it is equally viable to consider $\cH$-treewidth for other choices of $\cH$. Naturally, one should aim at graph classes where problems of interest become tractable, but it is also important to make sure that a (nice) $\cH$-tree decomposition can be computed efficiently (i.e., one needs to obtain analogues to our Lemma~\ref{lem:computing}). Examples of graph classes that may be explored in this context include split graphs, interval graphs, and more generally graphs of bounded mim-width~\cite{JaffkeKT18}. 


\paragraph{Acknowledgments.} Robert Ganian and Thekla Hamm acknowledge support by the Austrian Science Fund (FWF, project P31336). Robert Ganian is also affiliated with FI MUNI, Czech Republic.

\bibliographystyle{abbrv}
\bibliography{literature}

\end{document}